%% file: main.tex
\documentclass[
    english,
    parskip=half,
]
{scrartcl}

\input{tex/header}
\input{tex/tikz}

\input{tex/commands}

\input{adds/abbreviations}

\title{Quantum Software Ecosystem Design}

\author{Achim Basermann~\orcidlink{0000-0003-3637-3231}, 
        Michael Epping~\orcidlink{0000-0003-0950-6801}, 
        Benedikt Fauseweh~\orcidlink{0000-0002-4861-7101},
        Michael Felderer~\orcidlink{0000-0003-3818-4442},
        Elisabeth Lobe~\orcidlink{0000-0002-3473-8906},
        Melven Röhrig-Zöllner~\orcidlink{0000-0001-9851-5886},
        Gary Schmiedinghoff~\orcidlink{0000-0003-2259-7365},
        Peter K. Schuhmacher~\orcidlink{0000-0003-1232-4363},
        Yoshinta Setyawati~\orcidlink{0000-0003-3718-4491},
        Alexander Weinert~\orcidlink{0000-0001-8143-246X}}  

\affil{Institute of Software Technology, German Aerospace Center (DLR), 
       Linder Höhe, 51147 Köln, Germany, \email{elisabeth.lobe@dlr.de}}
\date{}

\usepackage[backend=biber]{biblatex}
\begin{document}

    \phantomsection\pdfbookmark[1]{Title}{title}
    \maketitle
    \vspace{-2\baselineskip}

    \abstract{%
        \pdfbookmark[2]{Abstract}{abstract}%
        \input{adds/abstract}    }

    \keywords{
        \pdfbookmark[2]{Keywords}{keywords}%
        Quantum Computing, Software Ecosystem, Hardware-Software Co-Design, Software Engineering
    }

\input{content/introduction}

\input{content/vision}

    \input{content/conceptual}

\input{content/technical}

\input{content/conclusion}

    \begin{acknowledgement}%
        \input{adds/acknowledgement}    \end{acknowledgement}

    \printbibliography

\end{document}

%% file: tex/header.tex
\usepackage{cmap}  
\usepackage[utf8]{inputenc}
\usepackage[T1]{fontenc}
\usepackage[english]{babel}

\usepackage{authblk} 
    
\usepackage{csquotes}
 \usepackage[breaklinks=true]{hyperref}

\usepackage{enumitem}
\usepackage{orcidlink}  
\usepackage{graphicx}  
\usepackage{quantikz}  
\usepackage{booktabs}
\usepackage{amsfonts}
\usepackage{todonotes}
\usepackage[abbreviations, hyperfirst=false]{glossaries-extra}

\usepackage{color}

\usepackage{mathtools}
    \mathtoolsset{showonlyrefs=true}

\AtBeginDocument{

}
\RedeclareSectionCommands[
  beforeskip=-.5\baselineskip,
  afterskip=.25\baselineskip
]{section,subsection,subsubsection}

\usepackage{bookmark}  
    \makeatletter
        \providecommand*{\toclevel@titlech}{0}
        \edef\toclevel@authorch{\the\numexpr\toclevel@titlech+1}
    \makeatother

\usepackage{xifthen}        
\usepackage{xstring}        

%% file: tex/tikz.tex
\usepackage{tikz}
\usepackage{tikzsymbols}
\usetikzlibrary{calc,shapes}
\tikzstyle{every picture}=[thick]

\newcommand{\VerConnector}[2]{\draw[thick,rounded corners,stealth-stealth] (#1) |- ($(#1) ! .5 ! (#2)$) -| (#2);}

\newsavebox{\userbox}
\savebox\userbox{
\begin{tikzpicture}[outer sep=0,inner sep=0]
  \draw (0,0) circle (.15);
  \draw (0,-.15) -- (0,-.5);
  \draw (0,-.5) -- (-.2,-.7);
  \draw (0,-.5) -- (.2,-.7);
  \draw (0,-.35) -- (-.2,-.15);
  \draw (0,-.35) -- (.2,-.15);
\end{tikzpicture}
}

\newsavebox{\desktopbox}
\savebox\desktopbox{
\begin{tikzpicture}[outer sep=0,inner sep=0]
  \draw[thick,rounded corners=.1cm,line width=.05cm,fill=white]
    (0,0) -- (.5,0) -- (.5,.7) -- (-.5,.7) -- (-.5,0) -- cycle;
    
  \draw[thick,rounded corners] (0,0) -- (0,-.2);
  \draw[thick,rounded corners] (-.3,-.18) -- (.3,-.18);
\end{tikzpicture}
}

\newsavebox{\documentbox}
\savebox{\documentbox}{\begin{tikzpicture}[scale=2]
	\draw[thick] (0,0) -- (0,-.5) -- (.4,-.5) -- (.4,-.1) -- (.3,0) -- cycle;
	\draw[thick] (.3,0) -- (.3,-.1) -- (.4,-.1);
	\foreach \y in {-.1,-.15,...,-.4} {
		\draw[thin] (.05,\y) -- (.35,\y);
	}
\end{tikzpicture}}

\newsavebox{\workstationbox}
\savebox\workstationbox{
\begin{tikzpicture}[thick]
  \draw (0,0) rectangle (.75,1);

  \draw (.1,.9) rectangle (.65,.8);
  \draw (.1,.8) rectangle (.65,.7);
  \draw (.1,.7) rectangle (.65,.6);

  \draw (.375,.2) circle (.08);
\end{tikzpicture}
}

\newsavebox{\serverbox}
\savebox\serverbox{
\begin{tikzpicture}[thick]
	\node[draw,rounded corners=.1cm,minimum height=.075cm,minimum width=1.2cm,fill=white] at (0,0) {};
	\node[circle,minimum size=.1cm,fill=black,inner sep=0] at (-.4cm,0cm) {};
	
	\node[draw,rounded corners=.1cm,minimum height=.075cm,minimum width=1.2cm,fill=white] at (0,.35) {};
	\node[circle,minimum size=.1cm,fill=black,inner sep=0] at (-.4cm,.35) {};
	
	\node[draw,rounded corners=.1cm,minimum height=.075cm,minimum width=1.2cm,fill=white] at (0,.7) {};
	\node[circle,minimum size=.1cm,fill=black,inner sep=0] at (-.4cm,.7) {};
\end{tikzpicture}
}

\newsavebox{\quantumbox}
\savebox\quantumbox{
\begin{tikzpicture}[outer sep=0,inner sep=0,scale=2]
  \draw[thick,fill=white] (0,0) circle (.05cm);
  \draw[thick] (0,0) ellipse (.2cm and .08cm);
  \draw[thick,rotate=60] (0,0) ellipse (.2cm and .08cm);
  \draw[thick,rotate=-60] (0,0) ellipse (.2cm and .08cm);
\end{tikzpicture}
}

\newsavebox{\circuitbox}
\savebox\circuitbox{
\begin{tikzpicture}[outer sep=0,inner sep=0,scale=.5]
  \draw (0,0) -- (1.5,0);
  \draw (0,-1) -- (1.5,-1);

  \draw[fill=black] (.75,0) circle (.1);
  \draw[] (.75,-1) circle (.2);
  \draw (.75,-1.2) -- (.75,0);

\end{tikzpicture}
}

\newsavebox{\programbox}
\savebox\programbox{
\begin{tikzpicture}[outer sep=0,inner sep=0]
  \draw[rounded corners] (1,0) -| (1.5,-1) -| (0,0) -- (1,0);
  \node at (.75,-.5) {$>$\_};
\end{tikzpicture}
}

\newsavebox{\browserbox}
\savebox\browserbox{
\begin{tikzpicture}[outer sep=0,inner sep=0]
  \draw[rounded corners] (1,0) -| (1.5,-1) -| (0,0) -- (1,0);
  \draw (0,-.1) -- (1.5,-.1);

  \draw (.75,-.55) circle (.3cm);
  \draw (.45,-.55) -- (1.05,-.55);
  \draw (.75,-.55) ellipse (.15cm and .3cm);
\end{tikzpicture}
}

\newsavebox{\cloudbox}
\savebox\cloudbox{
\begin{tikzpicture}[outer sep=0,inner sep=0]
  \node[draw,cloud,cloud puffs=12,minimum width=1cm,minimum height=.7cm] at (0,0) {};
\end{tikzpicture}
}

\newcommand{\gear}[6]{%
  (0:#2)
  \foreach \i [evaluate=\i as \n using {\i-1)*360/#1}] in {1,...,#1}{%
    arc (\n:\n+#4:#2) {[rounded corners=1pt] -- (\n+#4+#5:#3)
    arc (\n+#4+#5:\n+360/#1-#5:#3)} --  (\n+360/#1:#2)
  } -- cycle %
}

\newsavebox{\gearbox}
\savebox{\gearbox}{\begin{tikzpicture}
\begin{scope}[shift={(-.2cm, .2cm)}]
\draw[thick,fill=white] \gear{8}{.4cm}{.5cm}{22.5}{1}{0};
\end{scope}
\draw[thick,fill=white] \gear{8}{.4cm}{.5cm}{22.5}{1}{0};
\end{tikzpicture}}

%% file: tex/commands.tex
\newcommand{\bigo}{{\mathcal O}}

\providecommand*{\eg}{e.g.\@\xspace}
\providecommand*{\ie}{i.e.\@\xspace}
\providecommand*{\cf}{cf.\@\xspace}

\newcommand\createabbrv[3]{%
    \newabbreviation{#1}{#2}{#3}%
    \expandafter\newcommand\csname #2\endcsname{\gls*{#1}\xspace}%
}

\newcommand{\revision}[3]{%
    \todo{#1}%
    {\color{green}#2}%
    \ifthenelse{\isempty{#3}}{}{\todo[color=green]{#3}}%
}
\renewcommand{\revision}[3]{#2}  

\makeatletter
\def\provideenvironment{\@star@or@long\provide@environment}
\def\provide@environment#1{%
        \@ifundefined{#1}%
                {\def\reserved@a{\newenvironment{#1}}}%
                {\def\reserved@a{\renewenvironment{dummy@environ}}}%
        \reserved@a
}
\def\dummy@environ{}
\makeatother

\provideenvironment{keywords}{\textbf{Keywords: }}{}
\provideenvironment{acknowledgement}{\textbf{Acknowledgement: }}{}

\renewcommand{\abstract}[1]{\textbf{Abstract: }#1} 
\providecommand{\keywords}[1]{\begin{keywords}#1\end{keywords}}

\providecommand{\email}[1]{\href{mailto:#1}{\texttt{#1}}}

%% file: adds/abbreviations.tex
\createabbrv{dlr}{DLR}{German Aerospace Center}
\createabbrv{sc}{SC}{Institute for Software Technology}

\createabbrv{dqs}{DQS}{Digital quantum simulation}
\createabbrv{ide}{IDE}{integrated development environment}
\createabbrv{hpc}{HPC}{high-performance computing}
\createabbrv{qpu}{QPU}{quantum processing unit}
\createabbrv{qc}{QC}{Quantum computing}
\createabbrv{qa}{QA}{quantum annealing}
\createabbrv{aqc}{AQC}{adiabatic quantum computation}
\createabbrv{qite}{QITE}{quantum imaginary time evolution}
\createabbrv{nisq}{NISQ}{noisy intermediate-scale quantum}
\createabbrv{vqas}{VQAs}{variational quantum algorithms}
\createabbrv{vqes}{VQEs}{variational quantum eigensolvers}
\createabbrv{qaoa}{QAOA}{quantum approximate optimization algorithm}
\createabbrv{ir}{IR}{intermediate representations}
\createabbrv{tsc}{TSC}{traffic sequence charts}
\createabbrv{gpus}{GPUs}{graphics processing units}
\createabbrv{qubo}{QUBO}{quadratic unconstrained binary optimization}
\createabbrv{ml}{ML}{machine learning}
\createabbrv{ai}{AI}{artificial intellegence}
\createabbrv{cpu}{CPU}{central processing unit}
\createabbrv{sat}{SAT}{satisfiability}
\createabbrv{smt}{SMT}{Satisfiability Modulo Theory}
\createabbrv{qma}{QMA}{Quantum-Merlin-Arthur}
\createabbrv{sle}{SLE}{systems of linear equations}
\createabbrv{squids}{SQUIDs}{superconducting quantum interference devices}
\createabbrv{dag}{DAGs}{directed acyclic graphs}
\createabbrv{uml}{UML}{unified modeling language}

%% file: adds/abstract.tex
The rapid advancements in quantum computing necessitate a scientific and rigorous approach to the construction of a corresponding software ecosystem, 
a topic underexplored and primed for systematic investigation. 
This chapter takes an important step in this direction: It presents scientific considerations essential for building a quantum software ecosystem
that makes quantum computing available for scientific and industrial problem solving. 
Central to this discourse is the concept of hardware-software co-design, 
which fosters a bidirectional feedback loop from the application layer at the top of the software stack down to the hardware. 
This approach begins with compilers and low-level software that are specifically designed to align with the unique specifications and constraints of the quantum processor, 
proceeds with algorithms developed with a clear understanding of underlying hardware and computational model features, 
and extends to applications that effectively leverage the capabilities to achieve a quantum advantage.
We analyze the ecosystem from two critical perspectives: 
the conceptual view, focusing on theoretical foundations, 
and the technical infrastructure, addressing practical implementations around real quantum devices necessary for a functional ecosystem. 
This approach ensures that the focus is towards promising applications with optimized algorithm-circuit synergy, 
while ensuring a user-friendly design, an effective data management and an overall orchestration.
Our chapter thus offers a guide to the essential concepts and practical strategies 
necessary for developing a scientifically grounded quantum software ecosystem.

%% file: content/introduction.tex
\section{Introduction}
\label{sec:intro}

Over the past few decades, quantum computing has steadily garnered attention owing to its potentially transformative applications in various fields 
including cryptography~\cite{Shor97}, material science~\cite{Lloyd1073}, linear algebra~\cite{Harrow2009} and combinatorial optimization~\cite{kadowaki1998quantum}, among others. 
The possibility to vastly improve computational efficiencies in solving certain classes of problems, compared to classical computers, 
has driven significant interest and investment in quantum computing technologies from both the scientific community and the industry. 

In recent years the field reached a new level of maturity, characterized by the development of more stable qubit systems and increased gate fidelities~\cite{Byrd2023}.
The emergence of quantum hardware platforms from academia and industry has underlined the significant strides made in this direction, 
creating a foundation for more advanced research and practical explorations in quantum computing~\cite{Arute2019}. 
However, it must be acknowledged that while substantial, these advancements are but the precursors to a fully fault-tolerant quantum computing potential.

Despite the progress, the current era of \NISQ devices~\cite{Preskill2018quantum}, presents significant challenges, 
including limited qubit connectivity, low coherence times and gate cross-talk. 
Moreover, the reliable physical fabrication of these devices, especially on an industrial scale,
involves considerable hurdles: 
Ensuring the purity of materials, achieving the precise alignment of nanostructures, and maintaining the ultra-low temperatures necessary for operation present ongoing challenges.
Another problem is our limited understanding concerning the underlying principles of quantum algorithms, 
with a yet limited selection of algorithmic building blocks available, like the quantum Fourier transformation and the amplitude amplification.
The development of a diverse and comprehensive portfolio of high-level algorithms is central to advancing the quantum computing field. 

These factors naturally lead to the question: 
What is necessary to advance the field of quantum algorithms and how can we obtain meaningful results from these near-term quantum devices given the existing limitations? 
It is evident that, in the \NISQ era, the fruitful utilization of quantum devices necessitates approaches that can effectively navigate the noise and errors inherent to current hardware.

In answer to this central question, we propose the necessity of creating an ecosystem that uses an interdisciplinary approach grounded in the principle of hardware-software co-design.
This ecosystem requires the systematic development in software encompassing applications, algorithms, and compilers, 
and a robust technical infrastructure that is precisely aligned with the intricacies of existing and swiftly advancing quantum hardware. 
By establishing a framework where software development is intricately linked with hardware evolution, we aim to maximize the utility of quantum computing in its current \NISQ stage and beyond.
This approach does not exclude but rather complement hardware-agnostic abstractions
that allow for more generic software development independently of the specific hardware.

In our view, a quantum software ecosystem comprehends all aspects in and around software designed for quantum computers, 
\eg, novel quantum algorithms designed for specific devices, optimized compilers, pre- and post-processing tools for results from quantum computations, 
and the technical integration into existing \HPC environments.
It includes the whole path from user perspective over access to actual hardware 
and, reversely, from the embedded hardware access to the general availability for different end users. 

\medskip

In this review we first describe a potential vision, 
how such a quantum software ecosystem interfaces with the potential end users and with the quantum hardware in \autoref{sec:vision}. 
We then analyze the requirements for an efficient ecosystem from the conceptual view, focusing on abstract requirements and methods, in \autoref{sec:conceptual}. 
In \autoref{sec:technical} we are concerned with the technical implementation of such an ecosystem 
and finally in \autoref{sec:conclusion} we give a concise conclusion and an outlook for the potential of such a scientifically constructed software ecosystem.

%% file: content/vision.tex
\section{Quantum Computing Perspective}
\label{sec:vision}

Future applications of quantum algorithms have the potential to provide novel efficient solutions in various sectors. 
This includes breakthroughs in material science, such as new superconductors or ultrafast memory, 
solutions for industrial size planning problems, applications in cryptography or the design of new and more efficient drugs. 
In the following we describe how a quantum software ecosystem supports these aims, 
by interfacing the applications with the quantum devices in a comprehensive and user-centered way.

\subsection{Achieving the Vision Through the Quantum Software Ecosystem}
\label{sec:vision:perfect}

As the quantum computers continue to develop, it is plausible to predict a scenario 
where stakeholders, from academic researchers to industrial partners, gain access to quantum computational capabilities through cloud platforms. 
While such cloud access to quantum devices is already available for a limited number of platforms, 
the process is not yet streamlined and has various drawbacks due to the quantum device imperfections. 
However, such cloud-based access simplifies the challenges associated with operating and using quantum hardware, making it more feasible for a wider range of users.

At the heart of such a scenario, specialized quantum algorithms, devised by algorithmic developers, will be processed. 
In order to make these algorithms compatible with quantum hardware, specialized compilers, developed by experts in quantum software, will be crucial. 
These compilers will be responsible for translating high-level quantum logic into specific instructions, 
tailored for the distinct hardware platforms created by quantum hardware designers. 
Facilitating this process is the core responsibility of the quantum software ecosystem. 

Furthermore, an integral component of this ecosystem will be the integration of quantum computers with classical systems.
Fast embedded classical computers will process quantum-classical feedback algorithms within the coherence time of the quantum computer, especially those related to error correction. 
Additionally, \HPC frameworks will be instrumental for algorithms that use parameterized quantum circuits, 
as these often require intensive computations to optimize parameters in tandem with quantum processors.

Another component shaping this ecosystem is the principle of hardware-software co-design. 
In this paradigm, not only is software adapted to optimally exploit the capabilities of the underlying quantum hardware, 
but the design of future quantum processors is also influenced by application-driven requirements. 
This bidirectional feedback ensures that hardware evolution remains attuned to the practical needs and challenges posed by real-world quantum applications. 
By closely intertwining the development processes of both hardware and software, 
the co-design approach seeks to accelerate the maturation and optimization of the quantum computing landscape.

After the computations are completed, users will receive their results via the same cloud interface. 
This closed-loop system aims to streamline the process of quantum computing, from input to result retrieval, 
while maximizing efficiency and user accessibility. 
The sustainability and success of this vision are inherently tied to the collaborative effort 
between quantum algorithm developers, compiler specialists, hardware builders, software engineers and the users themselves.

\subsection{Interested Parties and Their Requirements} 
\label{sec:vision:stakeholders}

Research and development in \QC have accelerated dramatically in recent years. 
Due to its potential, efforts in \QC have attracted different parties. 
They are classified as primary and secondary stakeholders.
Primary stakeholders are stakeholders that directly contribute to the development of quantum computing as shown in \autoref{fig:stakeholder}.

\begin{enumerate}[leftmargin=*]
    \item End users:   
    End users are individuals or organizations from different fields that use or adopt \QC for various purposes, 
    \eg, to speed up simulations for electric car batteries, to predict financial risk in insurance companies or to optimize antenna patterns in radar technology. 
    They are influenced by design and functionality features provided by the \QC software researchers and the \QC hardware developers. 
    End users' expectations, values, and requirements must be considered to guarantee that the technology is effective and benefits them. 
    The end users may not know how to write the algorithm and formulate the problem as a quantum program, 
    but they can express it mathematically and are capable of post-processing the result of the computation as shown on the left panel of \autoref{fig:stakeholder}.

    \begin{figure}[t]
        \centering    
        \def\svgwidth{0.9\textwidth}
        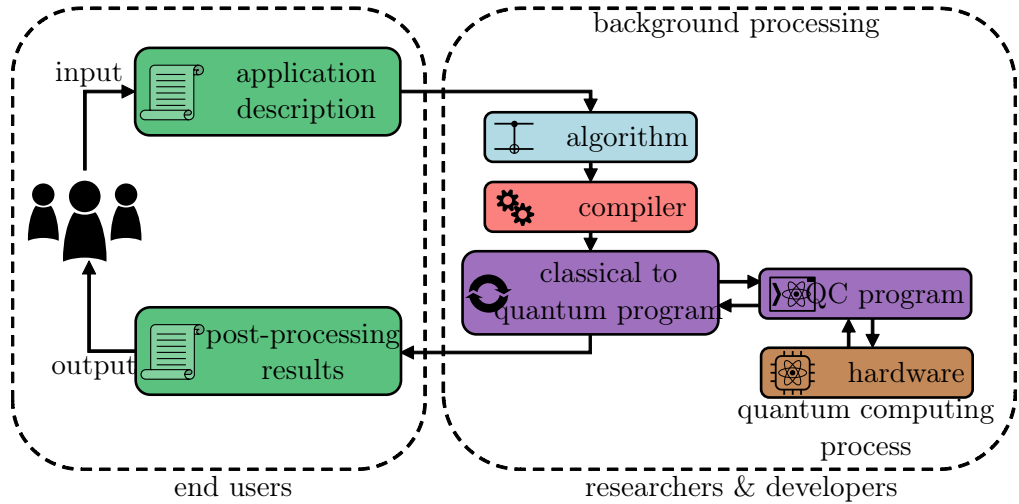
        \caption{Schematic diagram of the workflow and the stakeholders that directly use and develop quantum computing technologies.}
        \label{fig:stakeholder}
    \end{figure}

    \item Researchers and developers:
    They are individuals and organizations that are directly involved in the development of and research on \QC. 
    Currently, research institutes and universities are the primary sources of this group, 
    but also more and more large companies and start-ups participate in the development of \QC. 
    These vendors contribute significantly to the advancement of \QC, 
    for example, by developing hardware and software packages for industry and research institutions. 
    Their role is shown on the right panel of \autoref{fig:stakeholder} 
    and includes algorithmic problem descriptions, compilation, software, and hardware development, 
    such that the produced results can be post-processed and returned back to the end users. 
    Hence, their work influences the design and development of technology; 
    at the same time they must align with the goals of other stakeholders.
    \begin{enumerate}
        \item Software developers:
        These involve private companies or research institutions 
        that develop novel quantum algorithms, compilation schemes, and software interfaces between algorithmic solutions and hardware for \QC. 
        They also explore novel quantum computing architectures and investigate promising use cases for \QC. 
        Due to the noisy nature of current day devices, 
        the development has to take the low-level hardware properties into account to ensure optimal algorithm execution leading to unique design paradigms. 
        In this context, it is important to have a clear and precise understanding of the performance of components 
        and of the impact of physical quantum noise, which can be characterized by low-level benchmarks.
        \item Hardware designers:
        The development of physical quantum computers is crucial. 
        In many cases, hardware advancement is the bottleneck in the field of \QC. 
        Quantum computers are particularly sensitive to noise and errors caused by interactions with their surroundings. 
        This can lead to an accumulation of errors, lowering computation quality. 
        Thus, improving the fidelity of the hardware operations is critical, 
        even though noise can be tackled to some extent in software as well, see \autoref{sec:conceptual:errors}.
        Hardware manufacturers have a natural interest in making their devices available to a wide range of users.
        Some \QC hardware is developed by private companies which might restrict information about the implementation details and restrict access to low-level control features, 
        a fact that needs to be considered when developing software at the lower layers of the \QC stack. 
    \end{enumerate}
\end{enumerate}

Secondary stakeholders are interested parties who can influence the future of \QC but contribute indirectly to the workflow in \autoref{fig:stakeholder}. 
\begin{enumerate}[leftmargin=*]
    \item Suppliers:
    They provide the necessary equipment and spare parts to build \QC hardware. 
    These stakeholders should consider requests from researchers and developers, whose involvement can shape the design and availability of technology. 
    Semiconductor and chip manufacturers are a few examples of this stakeholder group. 
    The term `enabling technologies' is used in the context of \QC to denote the development of products and enhanced manufacturing techniques 
    that are not directly related to \QC itself but will facilitate breakthroughs in \QC and other fields.
    Therefore the suppliers play a crucial role in advancing the ecosystem. 

    \item Regulators and policymakers:
    They are responsible for the community's well-being and ensure that the developed technology boosts innovation. 
    These governmental entities are also responsible for ensuring that \QC aligns with society's values and needs, 
    for example by motivating the development of \QC to strengthen the economy and industrial advancement. 
    Hence, they create laws and regulations for the development and use of \QC. 
    In many situations, they provide state funding for research and development and encourage enterprises to foster the growth of~\QC.

    \item Investors:
    These are private funding sources that support research and development of \QC. 
    Investors are interested in the development of \QC and expect a return on investment in the future. 
    Investment in \QC has increased significantly from US\$93.5 million in 2015 to US\$1.02 billion in 2021 globally~\cite{pitchbook}. 
    Most investments are made for hardware, but there are also deals for software promising potential applications in the future.  

    \item Media:
    Media also play a significant role in the advancement of \QC technology. 
    They shape public opinion, hence raising awareness of \QC development and its impact on society. 
    They also convey the basic principles of this technology to the general public. 
    Not only the potential, but also the growth of research, technology, startups, and investment are communicated through media. 
\end{enumerate}

Only the collaborative effort between all of these stakeholders will enable quantum computing to establish as a well-founded technology, 
where the quantum software ecosystem should support the communication and form the baseline for further advancements.

%% file: figures/stakeholder.pdf_tex
\begingroup%
  \makeatletter%
  \providecommand\color[2][]{%
    \errmessage{(Inkscape) Color is used for the text in Inkscape, but the package 'color.sty' is not loaded}%
    \renewcommand\color[2][]{}%
  }%
  \providecommand\transparent[1]{%
    \errmessage{(Inkscape) Transparency is used (non-zero) for the text in Inkscape, but the package 'transparent.sty' is not loaded}%
    \renewcommand\transparent[1]{}%
  }%
  \providecommand\rotatebox[2]{#2}%
  \newcommand*\fsize{\dimexpr\f@size pt\relax}%
  \newcommand*\lineheight[1]{\fontsize{\fsize}{#1\fsize}\selectfont}%
  \ifx\svgwidth\undefined%
    \setlength{\unitlength}{566.92913386bp}%
    \ifx\svgscale\undefined%
      \relax%
    \else%
      \setlength{\unitlength}{\unitlength * \real{\svgscale}}%
    \fi%
  \else%
    \setlength{\unitlength}{\svgwidth}%
  \fi%
  \global\let\svgwidth\undefined%
  \global\let\svgscale\undefined%
  \makeatother%
  \begin{picture}(1,0.49534233)%
    \lineheight{1}%
    \setlength\tabcolsep{0pt}%
    \put(0,0){\includegraphics[width=\unitlength,page=1]{figures/stakeholder.pdf}}%
    \put(0.04376222,0.42128485){\color[rgb]{0,0,0}\makebox(0,0)[lt]{\lineheight{1.25}\smash{\begin{tabular}[t]{l}input\end{tabular}}}}%
    \put(0,0){\includegraphics[width=\unitlength,page=2]{figures/stakeholder.pdf}}%
    \put(0.67188865,0.28575972){\color[rgb]{0,0,0}\makebox(0,0)[rt]{\lineheight{1.25}\smash{\begin{tabular}[t]{r}compiler\end{tabular}}}}%
    \put(0.67394401,0.35649851){\color[rgb]{0,0,0}\makebox(0,0)[rt]{\lineheight{1.25}\smash{\begin{tabular}[t]{r}algorithm\end{tabular}}}}%
    \put(0,0){\includegraphics[width=\unitlength,page=3]{figures/stakeholder.pdf}}%
    \put(0.03993754,0.12713965){\color[rgb]{0,0,0}\makebox(0,0)[lt]{\lineheight{1.25}\smash{\begin{tabular}[t]{l}output\end{tabular}}}}%
    \put(0,0){\includegraphics[width=\unitlength,page=4]{figures/stakeholder.pdf}}%
    \put(0.72625808,0.00492117){\color[rgb]{0,0,0}\makebox(0,0)[t]{\lineheight{1.25}\smash{\begin{tabular}[t]{c}researchers \& developers\end{tabular}}}}%
    \put(0.72124717,0.47067675){\color[rgb]{0,0,0}\makebox(0,0)[t]{\lineheight{1.25}\smash{\begin{tabular}[t]{c}background processing\end{tabular}}}}%
    \put(0.85012771,0.08298239){\color[rgb]{0,0,0}\makebox(0,0)[t]{\lineheight{1.25}\smash{\begin{tabular}[t]{c}quantum computing\\process\end{tabular}}}}%
    \put(0.16201824,0.00497515){\color[rgb]{0,0,0}\makebox(0,0)[lt]{\lineheight{1.25}\smash{\begin{tabular}[t]{l}end users\end{tabular}}}}%
    \put(0,0){\includegraphics[width=\unitlength,page=5]{figures/stakeholder.pdf}}%
    \put(0.22314298,0.39319828){\color[rgb]{0,0,0}\makebox(0,0)[lt]{\lineheight{1.25}\smash{\begin{tabular}[t]{l}output\end{tabular}}}}%
    \put(0,0){\includegraphics[width=\unitlength,page=6]{figures/stakeholder.pdf}}%
    \put(0.59423469,0.21935042){\color[rgb]{0,0,0}\makebox(0,0)[t]{\lineheight{1.25}\smash{\begin{tabular}[t]{c}classical to\\quantum program\end{tabular}}}}%
    \put(0,0){\includegraphics[width=\unitlength,page=7]{figures/stakeholder.pdf}}%
    \put(0.94839476,0.19902628){\color[rgb]{0,0,0}\makebox(0,0)[rt]{\lineheight{1.25}\smash{\begin{tabular}[t]{r}QC program\end{tabular}}}}%
    \put(0.94850798,0.11914858){\color[rgb]{0,0,0}\makebox(0,0)[rt]{\lineheight{1.25}\smash{\begin{tabular}[t]{r}hardware\end{tabular}}}}%
    \put(0.29012507,0.15866534){\color[rgb]{0,0,0}\makebox(0,0)[t]{\lineheight{1.25}\smash{\begin{tabular}[t]{c}post-processing\\results\end{tabular}}}}%
    \put(0.2935012,0.42001152){\color[rgb]{0,0,0}\makebox(0,0)[t]{\lineheight{1.25}\smash{\begin{tabular}[t]{c}application\\description\end{tabular}}}}%
    \put(0,0){\includegraphics[width=\unitlength,page=8]{figures/stakeholder.pdf}}%
  \end{picture}%
\endgroup%

%% file: content/conceptual.tex
\section{Conceptual View}
\label{sec:conceptual}

\begin{figure}[t]
\centering
    \def\svgwidth{0.7\textwidth}
    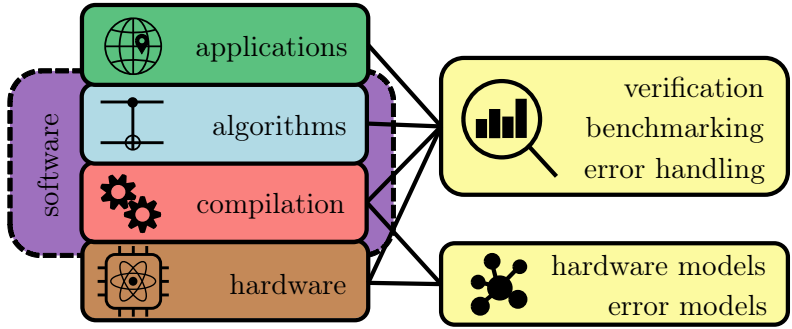
    \caption{Conceptual stack of the components necessary to solve problems using \QC.}
    \label{fig:conceptual_stack}
\end{figure}

In this section, we view the quantum software ecosystem from a more theoretical view,
focusing on conceptually important ideas and abstracted \QC concepts,
which are the main area of scientific research on \QC.
This conceptual view includes various topics, as shown in \autoref{fig:conceptual_stack}.

At the top of this ``stack'', \ie, on the side of the user, is the application or problem that needs to be solved,
and on the bottom of the stack  lies the hardware that executes the necessary \QC steps.
Those ends are connected by the software, including various algorithms and compilation schemes.
In order to attain the correct results, it is necessary to handle the noise-induced errors emerging during the computation,
which requires accurate error models for the hardware.
One major challenge is the verification of the various parts of this stack.
In the following, we look at each part of this stack and its role in the quantum software ecosystem.

\input{content/conceptual/qc_model}

\input{content/conceptual/hardware}

\input{content/conceptual/applications}

\input{content/conceptual/algorithms}

\input{content/conceptual/software}

\input{content/conceptual/compiler}

\input{content/conceptual/errors}

\input{content/conceptual/verification}

%% file: figures/conceptual_stack.pdf_tex
\begingroup%
  \makeatletter%
  \providecommand\color[2][]{%
    \errmessage{(Inkscape) Color is used for the text in Inkscape, but the package 'color.sty' is not loaded}%
    \renewcommand\color[2][]{}%
  }%
  \providecommand\transparent[1]{%
    \errmessage{(Inkscape) Transparency is used (non-zero) for the text in Inkscape, but the package 'transparent.sty' is not loaded}%
    \renewcommand\transparent[1]{}%
  }%
  \providecommand\rotatebox[2]{#2}%
  \newcommand*\fsize{\dimexpr\f@size pt\relax}%
  \newcommand*\lineheight[1]{\fontsize{\fsize}{#1\fsize}\selectfont}%
  \ifx\svgwidth\undefined%
    \setlength{\unitlength}{418.5982195bp}%
    \ifx\svgscale\undefined%
      \relax%
    \else%
      \setlength{\unitlength}{\unitlength * \real{\svgscale}}%
    \fi%
  \else%
    \setlength{\unitlength}{\svgwidth}%
  \fi%
  \global\let\svgwidth\undefined%
  \global\let\svgscale\undefined%
  \makeatother%
  \begin{picture}(1,0.41722809)%
    \lineheight{1}%
    \setlength\tabcolsep{0pt}%
    \put(0,0){\includegraphics[width=\unitlength,page=1]{figures/conceptual_stack.pdf}}%
    \put(0.06418849,0.20831595){\color[rgb]{0,0,0}\rotatebox{90}{\makebox(0,0)[t]{\lineheight{1.25}\smash{\begin{tabular}[t]{c}software\end{tabular}}}}}%
    \put(0,0){\includegraphics[width=\unitlength,page=2]{figures/conceptual_stack.pdf}}%
    \put(0.43290193,0.04240015){\color[rgb]{0,0,0}\makebox(0,0)[rt]{\lineheight{1.25}\smash{\begin{tabular}[t]{r}hardware\end{tabular}}}}%
    \put(0,0){\includegraphics[width=\unitlength,page=3]{figures/conceptual_stack.pdf}}%
    \put(0.4311982,0.14615975){\color[rgb]{0,0,0}\makebox(0,0)[rt]{\lineheight{1.25}\smash{\begin{tabular}[t]{r}compilation\end{tabular}}}}%
    \put(0,0){\includegraphics[width=\unitlength,page=4]{figures/conceptual_stack.pdf}}%
    \put(0.4329724,0.2476243){\color[rgb]{0,0,0}\makebox(0,0)[rt]{\lineheight{1.25}\smash{\begin{tabular}[t]{r}algorithms\end{tabular}}}}%
    \put(0,0){\includegraphics[width=\unitlength,page=5]{figures/conceptual_stack.pdf}}%
    \put(0.43430271,0.34882884){\color[rgb]{0,0,0}\makebox(0,0)[rt]{\lineheight{1.25}\smash{\begin{tabular}[t]{r}applications\end{tabular}}}}%
    \put(0,0){\includegraphics[width=\unitlength,page=6]{figures/conceptual_stack.pdf}}%
    \put(0.96944279,0.066593){\color[rgb]{0,0,0}\makebox(0,0)[rt]{\lineheight{1.25}\smash{\begin{tabular}[t]{r}hardware models\end{tabular}}}}%
    \put(0.97102732,0.01499492){\color[rgb]{0,0,0}\makebox(0,0)[rt]{\lineheight{1.25}\smash{\begin{tabular}[t]{r}error models\end{tabular}}}}%
    \put(0,0){\includegraphics[width=\unitlength,page=7]{figures/conceptual_stack.pdf}}%
    \put(0.96709814,0.2955369){\color[rgb]{0,0,0}\makebox(0,0)[rt]{\lineheight{1.25}\smash{\begin{tabular}[t]{r}verification\end{tabular}}}}%
    \put(0.96685263,0.24393877){\color[rgb]{0,0,0}\makebox(0,0)[rt]{\lineheight{1.25}\smash{\begin{tabular}[t]{r}benchmarking\end{tabular}}}}%
    \put(0.96810809,0.1923373){\color[rgb]{0,0,0}\makebox(0,0)[rt]{\lineheight{1.25}\smash{\begin{tabular}[t]{r}error handling\end{tabular}}}}%
    \put(0,0){\includegraphics[width=\unitlength,page=8]{figures/conceptual_stack.pdf}}%
  \end{picture}%
\endgroup%

%% file: content/conceptual/qc_model.tex
\subsection{Computational Paradigms}
\label{sec:conceptual:hardware_model}

The development of a functional quantum computer is a central research goal these days.
There exist different paradigms how such a machine could look like even on a conceptual level.
In this section, we first review the basic principles of quantum mechanics on which all these quantum computing paradigms rely on.
Afterwards, we discuss the most prominent ones, namely the gate-based model and adiabatic quantum computation.
Finally, we briefly mention a few alternatives.

\subsubsection{Foundations of Quantum Computing}
\label{sec:conceptual:hardware_model:gate}

In this section, we outline the phenomenology that builds the foundation of \QC
without elucidating the rich mathematical framework of quantum mechanics that can be found in many textbooks~\cite{cohen1991quantum,nielsen00}.

A quantum bit, or qubit in short, is a direct generalization of a classical bit
with two additional, inherently quantum-mechanical properties: superposition, and entanglement with other qubits.
While a classical bit can only take one of the two states 0 and 1,
a qubit can be in a superposition of both at the same time. 
Mathematically, the state of a single qubit can be expressed as 
\begin{equation}
\label{superposition}
    |\psi\rangle=a|0\rangle+b|1\rangle,
\end{equation}
where $|0\rangle$ and $|1\rangle$ denote the computational basis states
written in Dirac notation that is convenient in quantum mechanics
and $a$ and $b$ are complex numbers with $|a|^2+|b|^2=1$.
The probability of measuring the state $|0\rangle$, \ie, a bit 0, is given by $|a|^2$
and analogously for $|1\rangle$ by $|b|^2$.
After measurement, the state of the qubit collapses to only the parts in agreement with the measurement outcome, 
\ie, $|\psi_0\rangle=|0\rangle$ or $|\psi_1\rangle=|1\rangle$.

\begin{figure}[b!]
    \centering
    \includegraphics[width=.35\textwidth]{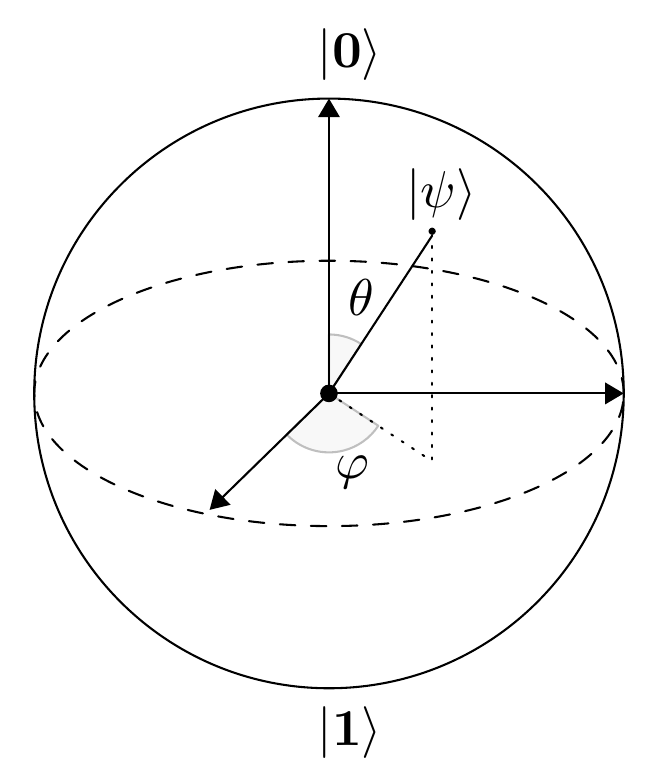}
    \caption{Visualization of an arbitrary qubit state called the Bloch sphere. 
             The computational basis states $|0\rangle$ and $|1\rangle$ are mapped to the north pole and the south pole respectively. 
             A general state $|\psi\rangle$ is fully determined by the angles $\theta$ and $\varphi$. 
             Any quantum gate on a single qubit corresponds to a rotation of the state on that sphere. 
             Graphic taken from~\cite{Schuhmacher2021}.}
    \label{fig:Bloch}
\end{figure}

Since $a$ and $b$ are complex numbers, they each contain a phase ($a=|a|e^{\text{i}\varphi_a}$).
In quantum mechanics, only the phase difference $\varphi=\varphi_b-\varphi_a$ is relevant,
hence the single-qubit state can be fully expressed by one probability and the relative phase, or equally by two angles.
Thus, any single-qubit state can be visualized as a unit vector
\begin{equation}
    |\psi\rangle=\begin{pmatrix}\sin\left(\theta\right)\cos\left(\varphi\right) \\ \sin\left(\theta\right)\sin\left(\varphi\right) \\ \cos\left(\theta\right) \end{pmatrix}
\end{equation}
on the Bloch sphere, which is depicted in \autoref{fig:Bloch}.
This visualization is also useful to understand the concept of the computational basis:
Any two opposite points on the Bloch sphere can be chosen as the computational basis states $|0\rangle$ and $|1\rangle$ 
and changing the basis is equivalent to rotating the qubit state. 

A superposition state needs to be initialized using classical information
and after performing a measurement collapses to one of these two states, \ie, back to a classical bit.
Therefore, the input and output is always restricted to classical bits,
but during the computation the full space of superpositions can be exploited.
This needs to be stressed: While a register of $N$ classical bits can describe one of $2^N$ different states at a time,
a $N$-qubit register can describe any state in a continuous region of a $2^N$-dimensional vector space.
As a consequence, qubits are tremendously more expressive than bits.
Since each measurement can change the qubit state $|\psi\rangle$, 
consecutive measurements of the same qubit in different bases do not yield additional information, 
unless one prepares $|\psi\rangle$ anew for each measurement.

The second important property of qubits, quantum entanglement,
is the ability of multiple qubits to interfere with one another
such that their probabilities become correlated in a way that is not possible for classical bits. 
For instance, two qubits can be entangled in the state $|\psi\rangle=a|00\rangle+b|11\rangle$.
When measuring the state of one the qubits, 
the result automatically determines the state of the other qubit in the same computational basis,
since \eg finding $|0\rangle$ for the first qubit collapses the full state to $|\psi_0\rangle=|00\rangle$. 

It is noteworthy that any computation on the full qubit state $|\psi\rangle$ acts on all superposed states at the same time, 
\eg on both $|00\rangle$ and $|11\rangle$.
This is utilized by many powerful quantum algorithms
that perform computations using precisely choreographed patterns of interference between superposition of bit strings,
which together with quantum entanglement realize the quantum computational efficiency. 
One needs to remember that measuring all qubits in a register collapses the carefully computed quantum state to a classical bit string,
so care must be taken to prepare the final quantum state in a way that maximizes the probability of measuring the bit string
that contains the relevant computational result.

Any natural or artificial quantum mechanical two-level system could in principle serve as a qubit,
making the number of possible realizations incredible large.
However, for fault tolerance a hardware platform needs at least to satisfy the DiVincenzo criteria~\cite{divincenzo2000physical}:
It is necessary to have  
\begin{enumerate}
    \item a scalable physical system with well characterized qubits,
    \item the ability to initialize the state of the qubits to a simple state,
    \item long relevant coherence times,
    \item an universal set of gates,
    \item and a qubit specific measurement capability.
\end{enumerate}
These qualitative criteria point out immediately why building a functional quantum computer remains a challenge up to date: 
On the one hand, satisfying criterion 3 requires to decouple the quantum system from any environmental disturbances. 
On the other hand, the criteria 2, 4 and 5 demand direct physical access to the system 
and therefore, it is necessary to couple it at least to its measurement apparatus and some control electronics. 
This ambivalence makes quantum computers inherently error prone. 
As of now, no quantum system exists that fulfills all criteria equivalently, 
but recent quantum hardware has reached a level of maturity that allows for small scale quantum computations. 
Platforms that have reached this level are dubbed \NISQ devices.

\subsubsection{Gate-Based Quantum Computing}
\label{sec:conceptual:hardware_model:gate_based}

In this section, we review the paradigm of gate-based quantum computing, 
which was the first quantum computing paradigm that has been proposed~\cite{nielsen00}. 
Here, a quantum gate denotes the analogue of a logical gate in classical computing. 
In the latter, there are only two possible gates on a single bit, namely the identity and the negation.
By contrast, any operation corresponding to a rotation on the Bloch sphere, shown in \autoref{fig:Bloch}, represents a valid quantum gate on a single qubit. 
Therefore, the set of valid quantum gates is uncountable even for that single qubit. 

In order to realize an actually useful quantum computer, it does not suffice to consider single-qubit rotations. 
Instead, we need a $N$-qubit register, and we need to be able to apply multi-qubit gates on any set of qubits. 
Fortuitously, it turns out to be sufficient to have access to just a single maximally-entangling two-qubit gate and to arbitrary single-qubit rotations to achieve universality~\cite{Sleator1995}. 
In other words, any quantum gate applied to the $N$-qubit register can be realized as a sequence of these elementary gates. 
There are multiple universal gate sets. In many cases the \QC hardware provides a basic set of gates, which ideally is universal.

One important consequence of quantum mechanical dynamics is that valid quantum gates must be unitary, 
\ie, the gate operations are represented by unitary matrices, which are reversible. 
Therefore, classical logic gates like the AND-gate, which has two input bits and one output bit, 
cannot be implemented directly on qubits without a second output qubit to ensure reversibility. 
Another consequence is that it is not possible to fully clone arbitrary qubit states, 
turning error correction by redundancy into a challenging prospect.

\begin{figure}[t!]
    \centering
    \begin{quantikz}
			\lstick{$\ket{\psi}$} & \qw & \ctrl{1} & \gate{H}  & \meter{} \vcw{2} & \\
			\lstick{$\ket{+}$} & \ctrl{1} & \targ  & \qw       &  \meter{}\vcw{1} & \\
			\lstick{$\ket{0}$}& \targ & \qw &  \qw  & \gate{X}    & \gate{Z} & \qw \rstick{$\ket{\psi}$}
		\end{quantikz}
    \caption{An example of a circuit diagram, the most common way to represent quantum programs today. 
             Horizontal lines correspond to qubits. 
             Gates are represented by special symbols or boxes with labels. 
             Double lines indicate classical information, which can represent results of the circuit.
             But they can also be used to condition the application of gates on measurement results, a technique called feed-forward.}
    \label{fig:circuitdiagram}
\end{figure}
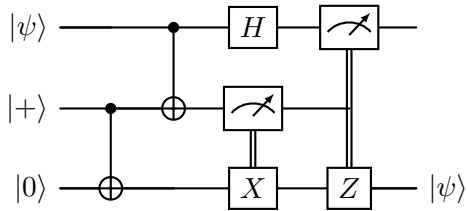

A sequence of quantum gates that solves a computational task composes a quantum algorithm. 
Quantum circuit diagrams have become established as a mode of representation, 
where the individual qubits usually correspond to horizontal lines on which gate operations are drawn (time runs from left to right)~\cite{nielsen00}. 
An example can be seen in \autoref{fig:circuitdiagram}.

\subsubsection{Adiabatic Quantum Computation and Quantum Annealing}
\label{sec:conceptual:hardware_model:annealing}

Around 2000, a new computational concept based on quantum mechanical principles was developed, 
the \AQC~\cite{farhi2000quantum}. 
The underlying adiabatic theorem is a fundamental result in quantum mechanics, 
originally formulated in~\cite{born1928beweis}.
The \AQC paradigm is different to the "conventional" quantum computing in the way 
that it does not provide a universal programmability straightforwardly in terms of implementing quantum gates to form quantum circuits. 
It rather represents a single algorithm whose input data can be varied. 
Nevertheless, the authors of~\cite{van2001powerful} and~\cite{aharonov2008adiabatic} have shown that \QC and \AQC are equivalent 
in the sense that each can efficiently simulate the other. 
We briefly summarize the main background of \AQC here, but for a more detailed review, 
we refer to~\cite{albash2018adiabatic}.

Given two related quantum systems, the rapid transfer from one to another might cause the system to change its state from their lowest-energy state, \ie, the ground state. 
By applying an adiabatic evolution process instead, that means a sufficiently slow transformation according to the adiabatic theorem, 
the system can however remain in its instantaneous ground state with high probability.  
By encoding a mathematical optimization problem in the target quantum system, 
where the energy states represent the feasible solutions, 
we could thus obtain the minimal solution of the problem. 

The first company who strives to built quantum systems based on \AQC and makes them commercially available is D-Wave Systems Inc. 
They implement the transverse field Ising model~\cite{Pfeuty,fauseweh2013}, established by Ernst Ising, 
using superconducting loops to form qubits in a quantum system~\cite{johnson2011quantum}. 
A current flow induces a magnetic flux in these loops, pointing either up or down or being in a superposition of both. 
Due to couplings of the loops by joints, the qubits interact with each other pairwise, 
where the strengths of the interactions can be adjusted with external magnetic fields. 
This way we can encode a quadratic function over binary variables, with linear and quadratic terms weighted according to the magnetic field strength. 
To find the solution of such a \QUBO problem is hard on classical computers. 
More precisely, its corresponding decision problem belongs to the class of NP-hard problems.
This also means it relates to a large number of other problems, 
which can easily be transferred into a \QUBO
and therefore solved with these machines, at least in theory. 

Although empirical studies like~\cite{junger2021quantum} provide hints that the output of the devices 
is in general close to the optimal solution, it is however not guaranteed to be achieved, 
nor is the success probability known in advance. 
Several physical restrictions prevent the realization of the theoretical concept of the adiabatic theorem, 
which only applies if ideal conditions prevail. 
One obstacle is, for instance, the shielding against environmental noise, which is never entirely achieved. 
Therefore, the term \QA has established, in reference to the classical heuristic simulated annealing, to distinct the theoretical concept from the heuristic process performed by the corresponding devices~\cite{mcgeoch2020theory}. 
In general, quantum annealing is repeated several times with the same configuration 
to obtain a sample set of solutions and from those the best one is extracted. 

\subsubsection{Others}
\label{sec:conceptual:hardware_model:other}

The gate-based model and quantum annealing are without question the leading quantum computing paradigms. 
However, there exist alternative paradigms that turn out to be computationally equivalent to these mainstream approaches. 
For example, a paradigm called one-way quantum computing is pursued in the context of photonic quantum computers~\cite{Raussendorf2001}. 
As photons hardly interact in nature, they can have enormous coherence times 
(one detects coherent photons from other stars regularly),
but it is a challenge to perform two-qubit gates between them for the same reason. 
In order to circumvent this issue, an elegant idea that relies on the Knill-Laflamme-Milburn proposal~\cite{Knill2001}, 
is to  prepare all entanglement non-deterministically first. 
If successful, then the computation is proceeded by measurements and single-qubit rotations only, 
\ie, by avoiding any further interaction~\cite{Browne2005}. 
However, functional one-way quantum computing has not been demonstrated yet.

Another universal approach for quantum computation are quantum random walks, or short quantum walks, 
a quantum mechanical analogue to the classical random walk~\cite{Aharonov1993,Kempe2003,Childs2003,Childs2009,Lovett2010}. 
They can either be discrete-time~\cite{Aharonov2001} or continuous-time~\cite{Farhi1998}, 
and they are studied in the context of machine learning~\cite{Schuld2014,Rebentrost2014} and photosynthesis~\cite{Mohseni2008b}. 
Both versions can again be extended to non-unitary evolution by a joint generalization of quantum and classical random walks, 
called quantum stochastic walks~\cite{Whitfield2010,Govia2017,Schuhmacher2021b}. 
In contrast to the completely coherent quantum walk, quantum stochastic walks give rise to a directed evolution.

%% file: content/conceptual/hardware.tex
\subsection{Hardware}
\label{sec:conceptual:hardware}

In 1936, Alan Turing proposed a conceptual blueprint for an universally programmable computer~\cite{turing1937computable}. 
This event became the child birth of modern computer science. 
However, as the direct physical implementation of the "Turing machine" would be impractical, 
a huge variety of different hardware platforms were used to realize different computational models. 
This early time of modern computer science came to a sudden end with the invention of the transistor~\cite{NobelPrize1956}. 
Since then, the development of classical computers relies on the same key building blocks but miniaturizing them.

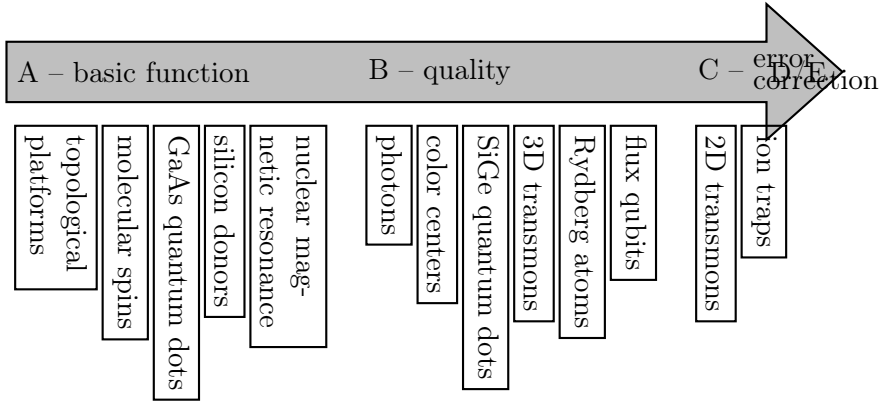
\begin{figure}[t]
    \centering
    \input{figures/hardware_levels}
    \caption{State of development of different hardware platforms according to~\cite{BSI_study}. 
             In this study, the platforms are classified into five different levels from satisfying the DiVincenco criteria (level~A), 
             demonstration of high fidelities (level~B) to the demonstration of quantum error correction (level~C). 
             The levels~D (execution of fault-tolerant operations) and~E (running fault-tolerant algorithms) have not been achieved by any platform so far.}
    \label{fig:BSI}
\end{figure}

In close analogy to these early days of classical computing, there exists a huge variety of candidates for quantum computing hardware 
-- the current status of quantum computer development resembles the construction of the Z3 by Konrad Zuse rather than building modern \HPC systems. 
A rather broad overview of hardware platforms, including a classification with respect to the state of development\footnote{Due to the status of the year 2020}, can be found in~\cite{BSI_study}. 
In the following, we focus on the most developed platforms according to this study, which are depicted in \autoref{fig:BSI}.

Generally speaking, there are two different classes of qubit candidates: 
natural quantum systems like neutral atoms, ions or photons~\cite{Knill2001,Blatt2008,Monroe2013,brandl2017quantum}, 
or artificial quantum systems like superconducting circuits or other solid state architectures~\cite{clarke2008superconducting,Kane1998,heinzel2008mesoscopic,Hayashi2003}. 
The state-of-the-art leading hardware platforms are based on trapped ions and planar transmons, 
the latter is a specific version of superconducting circuits. 
These platforms achieved the level of development C in \autoref{fig:BSI}, \ie, they allow for the demonstration of quantum error correction. 

Superconducting integrated circuits are viewed as one of the most promising hardware candidates~\cite{Acin_2018}. 
These circuits are put onto a chip that needs to be cooled to cryogenic temperatures, \ie, a few tens of mK, 
and they are controlled with electromagnetic fields in the microwave range. 
Even for this specific architecture, there is a variety of different qubit designs. 
However, all these designs share the same key ingredient, namely the Josephson junction~\cite{josephson1962possible}. 
This is a nonlinear element leading to a non-equidistant energy spectrum of the circuit. 
This property is crucial to address two quantum states as the computational states individually.

\begin{figure}[t]
    \begin{minipage}[t]{0.48\textwidth}
        \centering
        \includegraphics[height=3.5cm]{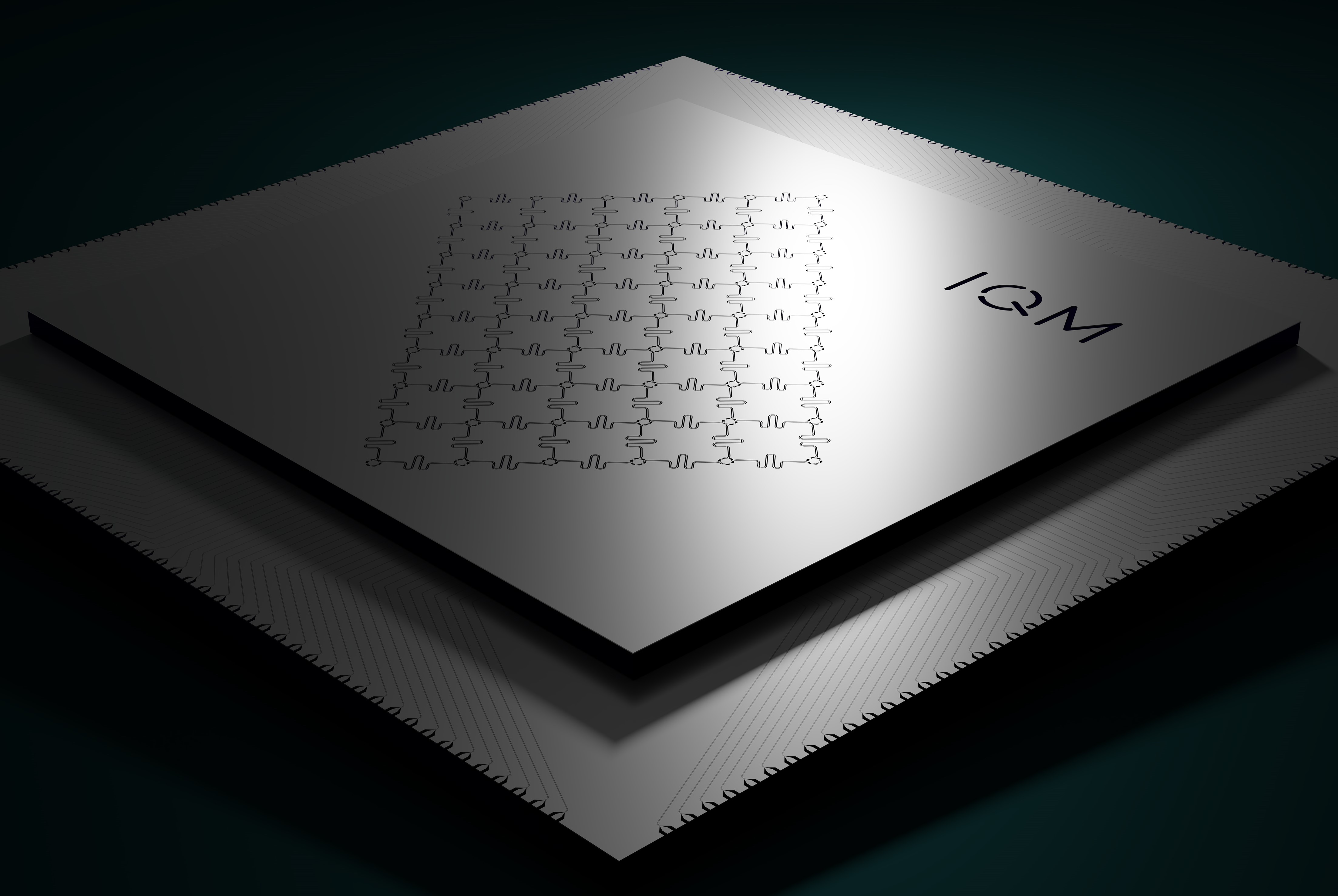}
        \caption{Sketch of the KQCircuits chip design by the company IQM Quantum Computers (courtesy of IQM Quantum Computers).}
        \label{fig:IQM}
    \end{minipage}\hfill
    \begin{minipage}[t]{0.48\textwidth}
        \centering
        \includegraphics[height=3.5cm]{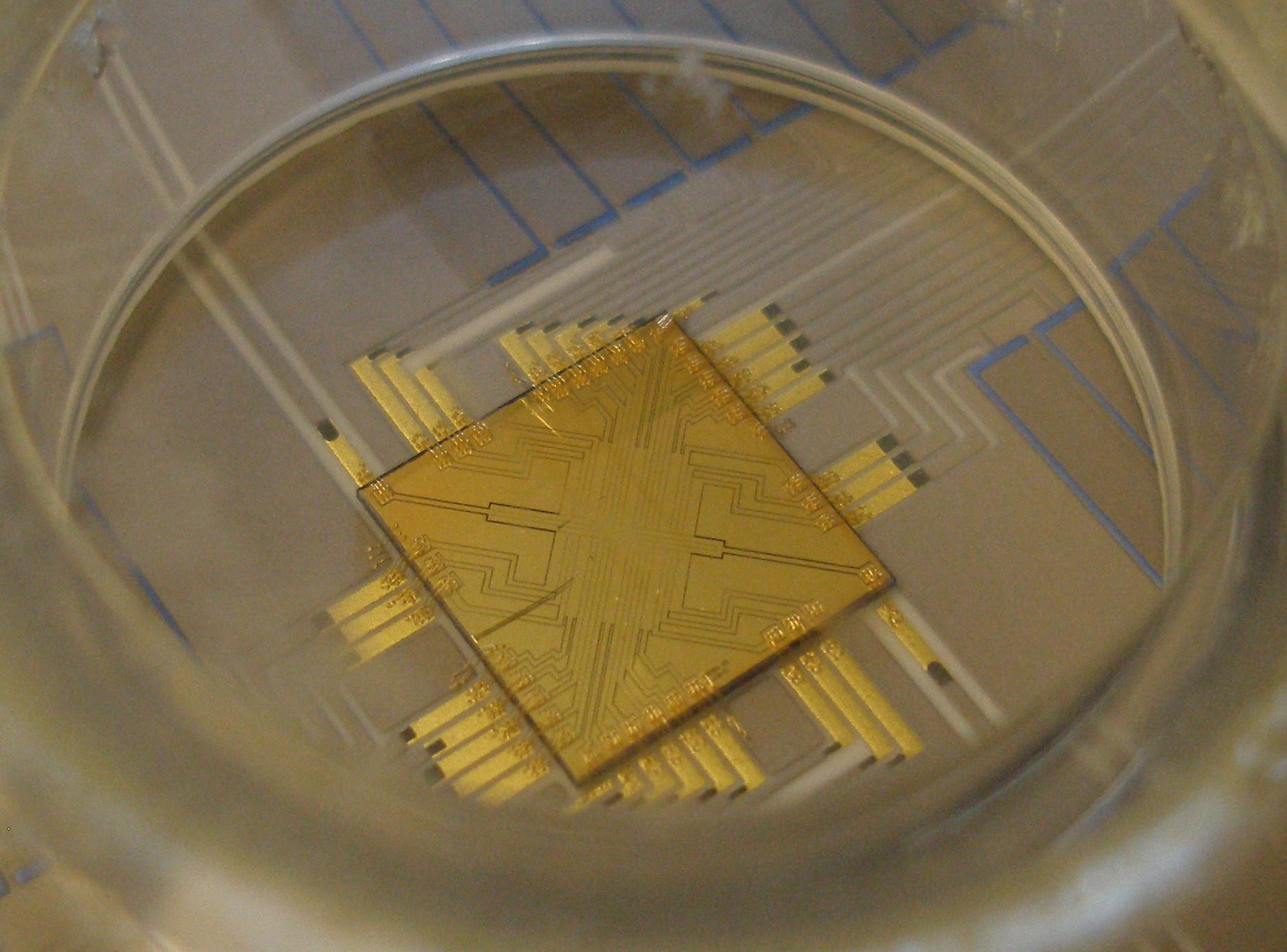}
        \caption{Photograph of the surface trap chip design by the company eleQtron (courtesy of eleQtron).}
        \label{fig:eleQtron}
    \end{minipage}
\end{figure}

There are two mainstream types of superconducting qubits, \ie, charge qubit~\cite{Bladh_2005,Vion2002,koch2007charge} and flux qubit~\cite{mooij1999josephson,van2000quantum,Pop2014} derived designs. 
To date, the primary representative of charge derived qubits is the planar transmon, 
due to its suppressed sensitivity against charge noise at the cost of small anharmonicities in the level splittings~\cite{koch2007charge,Houck2009}. 
It operates at a sweet spot with rather long coherence times and a good reproducability of the qubits. 
The main benefit of planar transmons is their rather straightforward scaling in qubit numbers; 
the challenge here is to maintain the controllability of the individual qubits and to keep high fidelity operations when scaling up. 
Transmons are typically considered for implementing gate-based quantum computing.
One draft of a corresponding chip is shown in \autoref{fig:IQM}, where the planar transmons are arranged in a two-dimensional square lattice with nearest-neighbour-interactions. 
Control and readout lines are connected to the qubits from below.

Flux qubits consist of superconducting loops that are interrupted by an (effectively) odd number of Josephson junctions. 
Their computational states are encoded in the magnetic fluxes that are induced by clockwise and anticlockwise circulating currents. 
By design, they share a lot of similarities to \SQUIDs~\cite{clarke2006squid}. 
Flux qubits can be coupled easily via mutual induction with coupling constants up to ultra-strong coupling if needed. 
This makes them an auspicious candidate for quantum annealing, and possibly for specific quantum simulation applications. 
In comparison to planar transmons, flux qubits are easier to couple, but it is harder to reproduce them reliably.

Apart from technical challenges, one of the main drawbacks of superconducting qubits is their limited connectivity: 
Only nearest neighbours are directly coupled and hence, two-qubit-gates can only applied between them directly. 
If a gate-based quantum algorithm needs gates between qubits that are not physically connected, 
one needs to perform the desired logical gate by swapping the qubit state through the intermediate qubits. 
This process produces a serious overhead in circuit depth. 
For quantum annealing, the limited connectivity becomes even more serious, 
because general optimization problems require strongly connected problem Hamiltonians. 
Therefore, embedding the desired problem Hamiltonian on the actual hardware becomes a non-trivial task~\cite{lobe2022diss}. 
Moreover, as superconducting qubits are artificially made, every single qubit has slightly different parameters than the others, 
an issue that needs to be tackled by optimal control theory~\cite{Motzoi2009}.

\medskip

With up to about 20 qubits, the best performing quantum computer is a chain of isotopically pure ions in a linear Paul trap\footnote{Named after Nobel laureate Wolfgang Paul.}. 
The ions are trapped in ultrahigh vacuum using electromagnetic fields in a quadrupole geometry such that they form a one-dimensional crystal~\cite{PaulSteinwedel,Paul1990}. 
No cryogenics are needed, the trap operates at room temperature. 
In contrast to superconducting qubits, the ions in the trap are coupled via the long-range Coulomb interaction, 
leading to a natural all-to-all connectivity of the qubits. 
In comparison to other hardware platforms, the relevant coherence times are high, and the gate quality is excellent.

Unfortunately, the design of the linear Paul trap does not allow for a scaling to large qubit numbers for two reasons: 
On the one hand, adding more and more ions into the trap deforms their arrangement; 
the ions start to form two-dimensional structures instead of a well-controlled chain. 
On the other hand, an effect called frequency crowding becomes more and more dominant, such that the system becomes uncontrollable~\cite{Kielpinski2002}. 
Therefore, the main challenge for trap ion based quantum computing is the scaling to larger qubit numbers. 
One ansatz is to combine several linear Paul traps via photonic links~\cite{Monroe2014}. 
Here, the quantum information needs to be converted from the ions in the trap to photons that are transmitted through a fiber, 
and then it is converted back to the ions in another trap. 
This process makes quantum computing with trapped ions enormously slow, because every single conversion only succeeds with limited probability. 
A different strategy is to use two-dimensional surface traps instead of linear Paul traps~\cite{Lekitsch2017}. 
Here, the second dimension is used to shuttle the ions during the computation to different zones on the chip, 
depending on their current purpose (performing a gate, readout etc). 
In the gate zone, the surface trap mimics the linear Paul trap with its advantages locally. 
A photograph of such a surface trap is shown in \autoref{fig:eleQtron}. 
However, surface traps were not able to demonstrate the same quality as linear Paul traps yet.

In this section, we discussed the benefits and drawbacks of the furthest developed hardware platforms to date, namely superconducting circuits and ion traps. 
However, as the field develops rapidly, other platforms may take over in the future. 
But even in this case, the substantial challenges to build functional quantum computers will probably remain during the next decades~\cite{Preskill2018quantum}. 
Therefore, any near-term quantum software ecosystem needs to incorporate the specific hardware restrictions that are present or that are expected to remain in the near future. 
For example, one requires additional compilation techniques to run a desired quantum algorithm on a superconducting qubit platform due to its limited connectivity, 
as on ion trap platforms with natural all-to-all connectivity. 
Conversely, if a given quantum algorithm can be easily embedded on the connectivity graph of the superconducting chip, 
then this platform might be preferential because of the larger qubit numbers that can be achieved. 
In the long term, as soon as universal fault-tolerant quantum computers are realized, 
the necessity to keep track of the specific hardware limitations by designing a quantum software ecosystem will become less and less important.

%% file: figures/hardware_levels.tex
\begin{tikzpicture}                                                                                                      
    \tikzstyle{tec} = [rotate=-90, anchor=south west, draw]
    
    \def\arrowlength{10}
    \def\arrowwidth{0.8}
    \def\diffsmall{0.05}
    \def\difflarge{0.5}
    
    \draw[fill=gray!50] (0, 0.5*\arrowwidth) -- (\arrowlength, 0.5*\arrowwidth) -- (\arrowlength, 0.5*\arrowwidth + 0.5) -- ($(\arrowlength, 0) + (1, 0)$) 
                     -- (\arrowlength, -0.5*\arrowwidth - 0.5) -- (\arrowlength, -0.5*\arrowwidth) -- (0, -0.5*\arrowwidth) -- cycle;
    
    \node[tec, text width=5em] (tec01) at (0.1,-0.7) {topological platforms};
    \node[tec] (tec02) at ($(tec01.north west) + (\diffsmall, 0)$) {molecular spins};
    \node[tec] (tec03) at ($(tec02.north west) + (\diffsmall, 0)$) {GaAs quantum dots};
    \node[tec] (tec04) at ($(tec03.north west) + (\diffsmall, 0)$) {silicon donors};
    \node[tec, text width=7em] (tec05) at ($(tec04.north west) + (\diffsmall, 0)$) {nuclear magnetic resonance};
    \node[tec] (tec06) at ($(tec05.north west) + (\difflarge, 0)$) {photons};
    \node[tec] (tec07) at ($(tec06.north west) + (\diffsmall, 0)$) {color centers};
    \node[tec] (tec08) at ($(tec07.north west) + (\diffsmall, 0)$) {SiGe quantum dots};
    \node[tec] (tec09) at ($(tec08.north west) + (\diffsmall, 0)$) {3D transmons};
    \node[tec] (tec10) at ($(tec09.north west) + (\diffsmall, 0)$) {Rydberg atoms};
    \node[tec] (tec11) at ($(tec10.north west) + (\diffsmall, 0)$) {flux qubits};
    \node[tec] (tec12) at ($(tec11.north west) + (\difflarge, 0)$) {2D transmons};
    \node[tec] (tec13) at ($(tec12.north west) + (\diffsmall, 0)$) {ion traps};
    
    \node[anchor=west] (A) at ($(tec01.south west) + (-0.1, 0.7)$) {A -- basic function};
    \node[anchor=west] (B) at ($(tec06.south west) + (-0.1, 0.7)$) {B -- quality};
    \node[anchor=west] (C) at ($(tec12.south west) + (-0.1, 0.7)$) {C -- \begin{minipage}{3cm}error \\[-0.4\baselineskip] correction\end{minipage}};
    \node[anchor=west] (DE) at (\arrowlength - 0.1, 0) {D\smash{/}E};
\end{tikzpicture}

%% file: content/conceptual/applications.tex
\subsection{Applications}
\label{sec:conceptual:applications}

Quantum computers have enabled advancements in a range of applications, 
starting with well-established domains such as database search and factorization using Grover's and Shor's Algorithms. 
These have a proven potential in enhancing search capabilities and disrupting traditional cryptographic methods, respectively, 
but require a level of fault tolerance not yet reached on quantum devices.

Beyond these utilities, quantum machine learning is emerging as a noteworthy area of application~\cite{Biamonte2017,Saggio2021}, 
enabling advancements in categorization, learning tasks, and the solution of partial differential equations. 
However, it is on the intermediate timeline where quantum simulation and optimization are drawing heightened attention. 
Quantum simulation facilitates the study of quantum systems, promising more accurate modeling of atomic and chemical processes, 
with applications in material science, quantum chemistry and drug design. 
In parallel, quantum optimization provides avenues for solving complex problems more efficiently, 
finding its relevance in logistics, finance, and more.

In the forthcoming sections, we narrow our focus on quantum simulation and optimization, 
as these represent the realms where quantum computing is expected to offer significant advantages in the near-term.

\subsubsection{Simulation}
\label{sec:conceptual:applications:sim}

\DQS represents a notable application for future quantum computers, focusing on simulating quantum systems with universal quantum computers. 
Richard P. Feynman originally suggested this application~\cite{feynman2018simulating}, later formalized by Lloyd~\cite{Lloyd1073}. 
\DQS is of particular significance for studying quantum materials like superconductors and topological insulators, which prove challenging for classical simulations.

Emergence, described as the rise of new system properties from the fundamental interactions of its components, 
has been evident in quantum phases and is directly connected to the existence of strong quantum fluctuations and entanglement. 
Traditionally, the examination of such phenomena relied on resource-intensive experiments, 
which explored only a limited range of parameters, including material composition and external electromagnetic fields. 
Theoretical modeling and simulation can significantly conserve resources and is pivotal for advancing material science. 
Yet, simulations of quantum models on conventional computers face challenges due to the exponential scaling with system size.
Classical simulations on modern \HPC hardware are capable to describe non-equilibrium dynamics 
in quantum dots~\cite{fauseweh2017efficient}, of 1D quantum systems~\cite{paeckel2020detecting}, as well as 2D systems~\cite{schwarz2020classification, fauseweh2020},
but with strong limitations in the simulatable system size.

\DQS employs quantum computers to efficiently simulate quantum systems. 
However, the current state of \DQS struggles to match the capabilities of conventional \HPC. 
Advancements in the present \NISQ hardware require innovative quantum algorithms like the \VQEs~\cite{Peruzzo2014}, 
which capitalizes on the increased expressiveness of quantum computers~\cite{fauseweh2023quantum}.
Ongoing research is centered on assessing the strengths and weaknesses of various hardware platforms concerning their potential \DQS applications~\cite{fauseweh2021digital}.

\subsubsection{Optimization}
\label{sec:conceptual:applications:opti}

Optimization problems appear in all fields where resources are limited, 
for instance in engineering, economics, computer science, and lots of others.
The development of efficient solution methods and answering the question whether those actually exist 
is the essential part of the research in mathematical optimization and complexity theory. 
A very important and well-studied class of problems are the NP-hard ones, 
which are, loosely speaking, those problems that cannot be solved efficiently using classical computation.
By simply increasing the computational resources of classical computers, this situation cannot be relaxed.
This naturally calls for the exploration of different, more powerful computational models. 
And the hope is that quantum computation steps into the breach due to properties of superposition, entanglement and quantum parallelism.

As explained in \autoref{sec:conceptual:hardware_model:annealing}, quantum annealing is a tailored method to solve discrete optimization problems. 
Several studies have shown the practical feasibility of this approach in different research areas, 
\eg, for the optimization of flight routes~\cite{stollenwerk2019quantum}, flight gate assignments~\cite{stollenwerk2019flight} and satellite scheduling~\cite{stollenwerk2021agile}. 
However, due to their heuristic nature, the actual practical advantage of the quantum annealers over dedicated classical approaches, 
including approximation algorithms and heuristics, is still under discussion.

Besides the optimization-tailored \QA, also algorithms for the gate-based quantum computing concept have been developed, 
like \QAOA or Grover search, which we elaborate in the next section.
However, due to currently too limited available resources, 
their performance on interesting industrial applications still needs to be evaluated in the future~\cite{Misra2023}. 
To investigate the capabilities of all such approaches systematically, they need to be integrated into a full software environment 
that allows to quickly formulate different applications and to benchmark the results of the quantum devices against several classical approaches.

%% file: content/conceptual/algorithms.tex
\subsection{Algorithms}
\label{sec:conceptual:algorithms}

The application cases described in \autoref{sec:conceptual:applications} can also, in principle, be solved on classical computers. 
In order to gain a speedup over these classical approaches by using quantum computers, efficient quantum algorithms are necessary.
While many promising algorithms already exist, there is active work on expanding the existing toolbox.
A quantum software ecosystem must provide a library of algorithms that end users can access
and must also support the development of new algorithms for domain and quantum experts.

The development of novel quantum algorithms faces two main challenges: 
Currently, there is much less experience in realizing quantum algorithms as software than for classical algorithms
and, in order for quantum algorithms to be viable, they need to provide a significant asymptotic speedup over existing classical algorithms. 
The following section provides a selection of important quantum algorithms, many of which provide super-polynomial speedup. 
A more extensive overview can be found in~\cite{QuantumAlgoZoo}.

\subsubsection{Powerful Algorithms for Fault-Tolerant Devices}
\label{sec:conceptual:algorithms:fault_tolerant}

\autoref{tab:SC_Algorithms_Overview:Fault_Tolerant} lists some of the most promising quantum algorithms~\cite{Montanaro2016},
which are expected to provide a quantum advantage on fully fault-tolerant \QC.
One such algorithm is Shor's algorithm for the prime factorization of large integers with super-polynomial speedup compared to the classical counterpart.
Shor's algorithm is based on the Quantum Fourier transformation and connected to the more general class of hidden subgroup problems,
which include \eg discrete logarithms and Gauss sums.
Grover's algorithm searches through an unsorted list with a polynomial speedup.
The quantum phase estimation algorithm approximates eigenvalues of a given Hamiltonian. 
Furthermore, a quantum computer can efficiently perform quantum time evolutions
and \SLE can be solved with the algorithm by Harrow, Hassidim and Lloyd~\cite{Harrow2009}.
A variety of other quantum algorithms, 
such as the Deutsch-Jozsa algorithm~\cite{deutsch1992rapid}, the Bernstein-Vazirani algorithm~\cite{bernstein1997quantum} and the Simon algorithm~\cite{simon1997on}, 
have been found as well, but won't be discussed here in detail.

\begin{table}[b!]
\centering
\newcommand{\nls}{\newline \hspace*{2mm}}
    \begin{tabular}{p{29mm}p{43mm}p{20mm}p{27mm}} \toprule
       Algorithm & Application Case & Complexity & Classical \nls Complexity \\ \midrule
       Shor      & Prime factorization of\nls integer with $N$ bits & $\bigo(N^2 \log N)$ & $\bigo(\exp( 1.9 N^{1/3} \nls ~~\times (\log N)^{2/3}))$ \\ 
       Quantum Fourier\nls Transform & Fourier transform with $N$ \nls amplitudes & $\bigo((\log N)^2)$ & $\bigo(N\log N)$ \\ 
       Grover & Unsorted search on $N$ \nls items & $\bigo(\sqrt{N})$ & $\bigo(N)$ \\ 
       Quantum Phase\nls Estimation & Eigenvalues of unitaries \nls up to error $\epsilon$ & $\bigo(1/\epsilon)$ & $\bigo(N^2)$ \\ 
       Harrow-Hassidim\nls-Lloyd & Solving \SLE with $N$ eqs. \nls and condition number $\kappa$ & $\bigo(\kappa^2 \log N)$ & $\mathcal{O}(\kappa N)$\\ \bottomrule
    \end{tabular}
    \caption{Examples of promising quantum algorithms for fault-tolerant \QC.}
    \label{tab:SC_Algorithms_Overview:Fault_Tolerant}
\end{table}

\subsubsection{Hybrid Algorithms for Noisy Intermediate Scale Devices}
\label{sec:conceptual:algorithms:nisq}

Fully fault-tolerant quantum computers are not expected to be built in the near future. 
Therefore, great effort is put into researching efficient algorithms for \NISQ devices, 
where the focus lies more on achieving quantum advantage over classical devices than on the best asymptotic performance. 
Many of these algorithms are heuristic and an asymptotic speedup is expected in special cases~\cite{TILLY20221}.
Some promising approaches in this area are listed in \autoref{tab:SC_Algorithms_Overview:NISQ}.

\begin{table}[t!]
\centering
\newcommand{\nls}{\newline \hspace*{2mm}}
    \begin{tabular}{p{15mm}p{46mm}p{50mm}p{12mm}} \toprule
       Algorithm & Application Case & Complexity & Classical \\ \midrule
         VQE & Eigenenergies and -states & Heuristic, often $\bigo(N^p)$ & $\bigo(e^N)$ \\
         QITE & Ground state preparation & For highly local Hamiltonians \nls $\bigo(N^p)$ & $\bigo(e^N)$ \\ 
         QAOA & Combinatorial optimization & Heuristic, potentially $\bigo(N^p)$ & $\bigo(e^N)$ \\
         \bottomrule
    \end{tabular}
    \caption{Examples of promising quantum algorithms for \NISQ devices.}
    \label{tab:SC_Algorithms_Overview:NISQ}
\end{table}

One general strategy to bring useful quantum algorithms on \NISQ devices is hybrid computation, 
where only the part of the problem that gains most from quantum hardware is solved on such, 
while the remaining problem is solved on a classical device. 
One example of this are \VQAs, most famously \VQEs~\cite{Cerezo2021}. 
The idea of \VQAs is to use a parameterized circuit on the quantum processor 
to prepare highly entangled states in the exponentially large Hilbert space and perform measurements on them.
The classical processor evaluates the measurement results and adapts the parameters of the quantum circuit in order to improve the result. 
For instance, \VQEs minimize the energy to find the ground state.
Various adaptions of this approach are being researched at the moment,
such as searching for excited states by optimizing the energy to be in a certain range or by enforcing orthogonality to the ground state.
Furthermore, ground states can be prepared efficiently for highly local Hamiltonians by using \QITE~\cite{McArdle2019}.

The \QAOA~\cite{Farhi2014} is used to solve combinatorial problems by encoding them as a Hamiltonian
with bit strings as representations of the possible solutions.
The \QAOA applies time evolution of a mixer Hamiltonian and problem Hamiltonian in alternation 
to find the bit string that minimizes the problem Hamiltonian expectation value.

A central challenge with performing these optimization algorithms in polynomial time is the risk of converging to local minima.
It is important to extend the scope of these algorithms and facilitate an infrastructure where a hybrid compiler, see \autoref{sec:conceptual:compiling},
can efficiently select which parts of a given problem to solve on the quantum device with quantum speedup.

%% file: content/conceptual/software.tex
\subsection{Software Engineering}
\label{sec:conceptual:se}

The goal of software engineering is the efficient development of high-quality software through scientific methods and precise processes. 
In this context, we understand software to be a structured collection of program code, documentation, quality assurance measures, artefacts 
and, where applicable, other data required to execute the programs. 
All software is written to perform specific tasks that can be described in the form of user stories: 
A user wants to achieve a goal with the software. 
The value of the software therefore lies in the efficient and reliable achievement of these goals.

Quantum software fits the above scheme just as well~\cite{serrano2022quantum}.
At this level of abstraction, the only difference is that quantum software contains parts that are executed on a quantum computer. 
As described above, quantum computers are particularly suitable for difficult problems, 
and the applications institutions, such as the \DLR\footnote{\url{www.dlr.de}},
are particularly interested in have a strong interdisciplinary character and will have a large scope. 
An efficient, structured approach and an integrated quality assurance strategy will therefore be essential in the near future.

In the following, we take a closer look at the aspects of software engineering,
where we recognise specific requirements of quantum computing or which, in our view, are particularly important in this context.

\subsubsection{Requirements}
\label{sec:conceptual:se:requirements}

A particular challenge in the development of software for quantum computers is the collection and specification of requirements~\cite{spoletini2023towards}.
It can differ significantly from classical software requirement engineering~\cite{yue2023towards}.

The first step is to describe the primary requirements. 
In our experience, this is done in collaboration with domain experts who often have little experience with quantum computers. 
Finding a common understanding of the problem to be solved is tedious, but always worthwhile. 
Subsequently, a precise mathematical formulation must be worked out that allows the mapping of the application to an existing quantum algorithm or the development of a new one.

The joint elaboration not only helps the software engineer to find a solution approach,
but also gives insights into quantum computing to a wider circle of interested people. 
This experience building within the organisation, but also within the ecosystem as a whole, is something we have recognised as its own value~\cite{Elevate,QCI}.

Secondary requirements arise from the primary ones, \eg requirements on the size of the system via the input data. 
The requirements must be considered together with the expected limitations of the hardware. 
A step that admittedly often leads to disillusionment and requires several iterations at this point. 
For instance at \DLR, there is a huge gap between the problem sizes that quantum computers can handle and the massive computing tasks that arise in engineering questions. 
However, we must and can already set the course for future advantages in our fields of application.
Despite or precisely because of the current hardware-related limitations, 
scalability must always be considered in quantum software development. 
There is great value today in demonstrating an algorithm that can solve small instances of difficult problems if it "only" needs to be scaled in the future, 
see~\cite{Vandersypen2001, Amico2019, Skosana2021} for the example of Shor's algorithm~\cite{Shor97}. 
In contrast, it seems questionable to implement a highly optimized algorithm that does not even theoretically scale to large instances.

\subsubsection{Software Design}
\label{sec:conceptual:se:design}

Software design is always about defining the architecture, components and their interfaces. 
In the design of quantum software, a dimension is added that is very important. 
It is necessary to decide which parts of the program are to be calculated on a conventional computer and which on a quantum computer. 
In this context, one also speaks of a \QPU, which can take over specific tasks. 
Not every task is well suited for a \QPU and it does not currently look as if quantum computers will completely replace conventional processors.

Once it has been determined what is to be computed where 
(which includes in particular the choice of a quantum algorithm as discussed above), 
a specification of the data exchanged between the conventional and quantum parts must be made.
A hardware-aware concept is required in order to feed data of a certain accuracy from a classical computer system reliably and accurately into a specific quantum circuit. 
Speed requirements here depend on the integration of the quantum hardware with the classical hardware and on the algorithm to be executed. 
Some hybrid algorithms require communication between the classical and the quantum system within the coherence time. 
The challenge here is to define abstract layers in the software design 
so that software solutions for reliable and accurate data communication between classical and quantum system are at least partially reusable.

A conceptual separation of software into hardware-specific and -agnostic parts increases the reusability of the software we develop. 
It is important to understand that, although we use very low-level methods to get the most out of our quantum computers, 
we aim to develop software and methods that are useful in the long term. 
Therefore, reusability is an important criterion.

Interfaces must be defined for the transfer of data. 
At present, there is practically no distinction between program code and data on the quantum computer. 
Input data is transferred via program code for preparing the data~\cite{Zhang2022}, which can be very hardware-specific, see~\cite{Cirac93, Wunderlich2009} for examples on ion traps. 
We expect that future abstractions will facilitate data transmission.

The development of suitable data types on quantum computers is still in its infancy. 
A lot of research and standardisation is still needed here. 
However, it is already apparent that, even for integers, the type of encoding has a major influence on the performance of quantum computers~\cite{Berwald2023}. 
Possible choices are amplitude encoding or basis encodings like binary encoding, one hot encoding, and domain wall encoding~\cite{Chancellor2019}.
More ways to encode classical data into quantum states are considered in the context of machine learning, see \eg~\cite{lloyd2020quantum}.
They affect the performance mainly due to the strong noise of current models, 
so any form of resource optimization can help a lot.

In many engineering applications, decimal fractions are of course required, 
which, depending on the required resolution, generate a very high resource requirement by today's standards (measured in number of qubits).
It can therefore be worthwhile to choose an algorithm that is formulated in data types that fit well with a quantum computer.

Finally, a good design process for quantum software includes simulations of the program 
and, if possible, test runs on available hardware. 
It allows challenges to be identified and the design to be adapted if necessary.
A rigid approach here is even more doomed to failure than in conventional software design.

\subsubsection{Models and Representation}
\label{sec:conceptual:se:models}

Let's take a look at current ways of representing quantum software, or rather program code for quantum computers. 
At the moment, mainly low-level descriptions are used. 
Even in most recent publications we are still on a level, that quantum algorithms are described via elementary gates and quantum circuits. 
Internally, these circuits can be represented as a list of gates, \DAGs, path integrals/phase polynomials or decision diagrams.
Low-level languages such as OpenQASM~\cite{OpenQASM}, cirq~\cite{Cirq} or qiskit~\cite{Qiskit} have become established as descriptions by a user and as interfaces between tools. 
Despite some attempts to create more high-level quantum programming languages, e.g. Q\#~\cite{Qsharp}, Silq~\cite{silq} or qrisp~\cite{qrisp}, none of these is currently widely used (for various reasons).
In the long term, however, there is no way around the introduction of more powerful language constructs in our view. 
It will be crucial that these find a natural way to represent the special capabilities of quantum computers. 
Although perhaps only years of programming experience will make natural programming languages for quantum computers possible, 
we want to support developments in this direction at an early stage.

In the context of compilers in particular, \IR are also introduced as an intermediate level between the abstraction layers of the programming language and the machine language. 
Examples are QIR~\cite{QIRSpec2021} and QSSA~\cite{Peduri2022}. 
The formulation of quantum-specific optimization steps on this level is a subject of current research,
which we will discuss in \autoref{sec:conceptual:compiling}.
We should also mention that other established tools of conventional software design are currently translated to and tried in the context of quantum computing, e.g. the \UML~\cite{PerezCastillo2022}.

\subsubsection{Software Testing}
\label{sec:conceptual:se:testing}

Software testing is part of the software development process that aims to ensure the quality and reliability of the software. 
There are different types of tests and common categories are unit tests (testing small components), integration tests, functional tests and acceptance tests (checking fulfilment of requirements). 
Tests are artefacts (code or instructions) that are executed automatically or manually. 
In contrast, verification relies on formal proofs which employ static code analysis, 
and benchmarking is concerned with the quantification of the performance of software and hardware. 
We look at verification and benchmarking in more detail in \autoref{sec:verification}. 
We emphasise that testing is about finding programming bugs, not hardware errors, whose treatment we discuss in \autoref{sec:conceptual:errors}. 
However, we investigate how the methods developed for handling hardware errors can also be adopted to testing.

Given the above definition of testing it is clear that future software for powerful quantum computers will also need to be tested. 
It is important to do basic preliminary work already now, before the hardware allows complex software to run. 
And research in this direction has indeed started~\cite{Usaola2020, Barrera2023, miranskyy2021testing, serrano2022quantum}. 
This ensures that the reliability of software does not become a bottleneck in future developments of \QC. 
It is particularly important, because in \QC the transition from low level circuits to high level programs mostly still lies ahead of us.
And testing is an exciting research topic in the field of quantum software engineering because quantum-specific phenomena have to be taken into account. 

It is obvious that facts like the no-cloning theorem \cite{Wootters1982} are obstacles in testing programs. 
Classical approaches that often use copying implicitly need to be adapted in order to apply them to quantum software. 
Also the fact that in general measurements in quantum theory disturb the observed system complicates state monitoring. 
This severely affects the possibilities for runtime tests on quantum computers, see \eg~\cite{li2020proq, Liu2020}, 
and further research in this direction will be necessary.

Furthermore what constitutes a typical error is quite different between classical and quantum programming. 
Due to the difference in the computational model there are even programming errors that do not meaningful in classical programming, 
\eg, when they affect only the phase of the state. 
It is therefore necessary to conduct studies on what bugs are typical in quantum programs~\cite{campos2021qbugs, zhao2021bugs4q}. 
Such studies can be very programming language specific, \ie, tailored towards Q\# \cite{Honarvar2020}. 
Only with knowledge about typical bugs it is possible to then develop good tests that detect as many of them as possible. 
A useful tool here is the creation of benchmark collections, as well as the automatic generation of test cases.

It is necessary to define tests that circumvent the above mentioned, quantum-specific challenges, and are still meaningful. 
And this leads to further research questions, such as the definition of meaningful measures for the significance of tests. 
Once more and more quantum software is written, guidelines for writing reliable code and informative tests which are based on the above research will be very useful.

%% file: content/conceptual/compiler.tex
\subsection{Compiling}
\label{sec:conceptual:compiling}
    
\subsubsection{Gate-Based Quantum Computing}
\label{sec:conceptual:gate}
    
Like conventional computers, quantum computers also implement a finite set of elementary basic operations, the gates already mentioned above.
Different sets of gates have become accepted for the description of quantum circuits~\cite{Gottesman98,Preskill98,nielsen00}.
If a gate set enables an efficient approximation of any unitary operation, we call it a universal gate set.
By efficient here we mean that the new length of the circuit scales polynomially with the original circuit length when switching from any other gate set.

On the one hand, convenient gate sets are used for the theoretical description of quantum algorithms.
These sets in general contain significantly more gates than necessary, are useful in the context of fault tolerance, and might also contain larger, undecomposed blocks.
On the other hand, each hardware platform implements different, sometimes very limited, gate sets.
A major restriction results, for example, from limited connectivity,
which means that the two-qubit operations provided are not possible for every pair of qubits.
However, some hardware platforms, in particular ion traps, provide native multi-qubit gates that allow to circumvent this issue, see also \autoref{sec:conceptual:hardware}.

The transition from one description of the quantum circuit to another is called \textit{transpiling}.
Specifically, the transition from a general unitary to a set of elementary gates is called \textit{synthesis}.
Both transitions are core tasks of a compiler.
Furthermore, the compiler, just like its conventional analog, has the task of customizing the output to the specific hardware as best as possible.
In summary, the requirements of a compiler are producing correct, efficient, and hardware-compatible output as explained in more detail below.

The typical compiler architecture can be divided into individual steps (passes),
which are connected in series as a pipeline where each step transforms the quantum circuit.
The best order is not obvious and passes can also be repeated at a later point in the compilation.
Typical transformation steps include:
\begin{itemize}
    \item \textbf{Synthesis}. Larger operations need to be decomposed into a universal set of basic gates.
          Small operations can be decomposed optimally w.r.t.\ a certain cost function,
          while synthesis of larger operations will not yield optimal solutions in general.
    \item \textbf{Routing}. The circuit needs to be rewritten in a way that contains only gates that are natively supported by the hardware.
          In particular, multi-qubit gates can only act on qubits that can interact physically.
          Even qubits that might not be part of a calculation can mediate the interaction.
    \item \textbf{Optimization}. The overall circuit can be optimized w.r.t.\ some cost function as well.
          Here the input and the output are both decomposed circuits.
          We discuss this point in more detail in the following.
\end{itemize}

Various objective functions for circuit optimization are conceivable and are used.
For example, the number of certain gates (\eg controlled-Not gate, T gate), the depth of the circuit (related but not identical to the runtime),
or the expected noise on the final state can be minimized.
Of course, one can also try to maximize algorithm-specific performance, \eg, the probability of success.
The problem of optimizing a circuit with respect to a particular objective is generally very difficult~\cite{Herr2017,botea2018,Amy2019},
such that there is no efficient algorithm to find the global minimum except for small circuits.
Some approaches are based on meet-in-the-middle~\cite{Amy2013} or \SAT solvers~\cite{Schneider2023}.
For larger circuits only heuristic algorithms are feasible, see \eg~\cite{Nam2018}.
For the optimization passes there are promising research approaches to transfer the conventionally established methods based on \IR to quantum compilers.

Deciding whether the result of the compilation is indeed efficient on actual devices is not obvious.
Since the global optimum is generally not known, one can only compare the result with other reference compilers.
However, this comparison depends strongly on the circuits.
It is important to use balanced benchmark suites, for example, the Arline Benchmark suite~\cite{kharkov2022arline}.
Further developments in this direction are foreseeable.

The development of compilers of hybrid programs has a major impact on the possibilities for optimization.
Such hybrid compilers are compilers that do not generate pure quantum circuits,
but executable code on conventional computers that contains calls to a \QPU~\cite{khalate2022llvmbased}.
This results in strong optimization potential
because the compiler can automatically decide whether the calculations are better to be computed on the \QPU or on a conventional computer.
Furthermore, it is even possible to apply hybrid simplification rules,
which, \eg, move individual operations from the quantum circuit to conventional pre- or post-processing~\cite{epping2022hybrid},
where they can be combined and simplified with established methods.
The goal is to leave only the essence of the quantum algorithm in the quantum circuit.
These hybrid simplification rules in particular can benefit from the established concept of \IR.

Another motivation for hybrid compilers is a closer coupling of the \CPU and the \QPU.
In particular, hybrid quantum algorithms such as \VQAs~\cite{Cerezo2021} benefit greatly from an efficient coupling of conventional and quantum systems.
Here, experience in GPU programming (\eg CUDA~\cite{cuda}) can be built upon.
In the future, abstract language constructs should simplify and unify the use of different hardware architectures.
\QPU and \CPU codes are developed in a common project folder, where calls to the \QPU are controlled via synchronous and asynchronous commands.
Efficient interfaces and protocols must be developed for uploading data and code to the \QPU and downloading measurement results.
The \QPU code may not only contain quantum operations
but also increasingly complex dynamic operations that directly process measurement results and influence subsequent quantum operations.
This feed-forward approach opens up exciting possibilities for new experiments,
and it is essential in the measurement-based model of quantum computation~\cite{Briegel2009}.
The close coupling of a CPU and a \QPU is flanked by development work
aimed at integrating quantum computers into \HPC environments, see~\cite{Lippert2022, HPCQC}.
The experience with these prototypes will influence the necessary standards.

As mentioned, we require the correctness of the compiler, \ie, a proof that the output corresponds in functionality to the original input, possibly in human-readable form.
However, it is known that the general equivalence test problem for quantum circuits is not efficiently solvable,
as it is in the class of QMA-complete problems~\cite{janzing2003identity},
which are, loosely speaking, those problems that are hard to solve for quantum computers.
This means that we have little hope of proving the correctness of the final result.
What we can do instead is to prove the correctness of the process.
The compiler is correct if it only applies correct transformations.
And for each individual transformation, it is possible to show correctness.
When we speak of heuristics in the compiler pipeline, we mean procedures that do not necessarily lead to improved circuits
but which nevertheless output a correct circuit in every case.

In addition to the described methodology, some approaches attempt to prove the equivalence of circuits.
Although they suffer from an exponential increase in resources (time or memory),
they can still deliver results for "simple" circuits.
We refer the interested reader here to the literature~\cite{Viamontes07,Burgholzer2020,Burgholzer2021}.

\subsubsection{Quantum Annealing}
\label{sec:conceptual:qa}

Although quantum annealing (\QA) is a different computational model and therefore poses its own challenges,
compiling in a certain sense is also needed here:
Quantum annealers can only process a very specific optimization problem,
in case of D-Wave, a restricted version of the Ising problem~\cite{lobe2022diss}.
Programming such devices essentially means providing the problem-defining parameters.
However, exemplary applications from industry and research, \cf~\autoref{sec:conceptual:applications:opti}, show that
there is in general no trivial way of obtaining these parameters.
Several transformation steps from the original problem formulation to the native one of the device are required.
A compiler handling these different abstraction layers would make the technology available for various users with different levels of expertise in \QA.

From a mathematical point of view, the step from an arbitrary discrete optimization problem  to a general Ising problem is solved
and can be done using a set of standard methods.
However, a complete software suite implementing this is not yet available.
Nevertheless, toolboxes like the D-Wave Ocean SDK~\cite{dwaveocean} or \texttt{quark}~\cite{quark} already support users with utility methods.
But further expansion of the software suites and conceptual advances are necessary.
For instance, the recent research on the reduction of combinatorial optimization problems has mainly focused on any kind of (polynomial) reduction
and not on the optimal one in a certain sense, \eg in the number of resulting variables,
which would be advantageous regarding the limited resources of current quantum computational devices.
Furthermore, the actually implemented Ising problem is not a general one but faces further restrictions,
such as a specific non-complete hardware graph and a limited parameter precision.
This causes the reformulation of the original problem to be a non-trivial task
and demands a ``compiler'' to implement all the necessary steps and hide the complexity from the application-focused users.

The two main transformation steps are the graph embedding and the parameter setting.
Unfortunately, the first step, the embedding of the original problem graph into the hardware graph, has appeared to be a computationally hard problem,
in particular, as hard as the problem D-Wave's annealers are capable of solving~\cite{lobe2021minor}.
Therefore, in practice, heuristic methods need to be applied to circumvent this bottleneck~\cite{cai2014practical,lobe2021embedding}.
In the second step, the hardware-native Ising problem has to be formulated based on the found embedding.
If this step is not done correctly, we will not be able to analyze the actual performance of the quantum device itself,
because the success probability might be suppressed due to a wrongly formulated problem.
Recently, a new formulation has been developed that provides an embedded Ising problem
which provably corresponds to the original problem and meanwhile optimizes its parameters with respect to the machine precision~\cite{lobe2023optimal}.
Based on this recent and future theoretical work,
the compilation software has to be steadily improved and extended,
and the full software ecosystem has to be able to adapt to these changes.

%% file: content/conceptual/errors.tex
\subsection{Error Handling}
\label{sec:conceptual:errors}

It is essential that errors caused by imperfect hardware are considered in the software stack, because the amplitudes and phases of the qubits are not discrete.
Additional steps or layers are necessary to protect the information against this unavoidable noise introduced by the hardware.
This section briefly sketches the main concepts in this field of research and provides references to more in-depth introductions.

We distinguish three categories of error-handling strategies. 
First, techniques that start directly at the hardware level and attempt to reduce the noise level~\cite{Lidar2014}. 
This includes, for example, dynamical decoupling~\cite{Viola1998}, where special control pulse sequences are used to eliminate the disturbing influence of the environment.
Second, techniques that encode the quantum information into subspaces 
that do not couple to the environment and are therefore not affected by decoherence introduced by the environment~\cite{Lidar2003}.
Third, taking the noise of the quantum computer into account for the compilation can lead to circuits that are less prone to noise. 
For example, we investigated which decompositions of a common multi-qubit gate introduce the least amount of noise~\cite{mueller2023coherent}. 

Furthermore, some post-processing steps on the classical measurement data are aimed at removing the noise~\cite{Cai2023quantum,endo2019a}. 
For example, zero-noise extrapolation~\cite{Temme2017} and readout error mitigation~\cite{beisel2022} have proven effective in some applications. 
The term error mitigation has come to refer to these types of techniques.
We emphasize that they do not avoid errors, but try to eliminate the errors afterwards.
Finally, methods on the error correction codes are aimed to suppress the noise to any degree~\cite{Gottesman2009introduction,bacon_2013,Roffe2019}. 
We explain error mitigation and error correction in more detail in the following texts.

\subsubsection{Error Models}
\label{sec:conceptual:errors:models}

Simple models for describing noise on quantum computers may only depend on a single or few parameters. 
They can be found in any textbook on quantum information, \eg~\cite{nielsen00}. 
The depolarising noise model is often used and can be interpreted in such a way 
that with a certain probability (the parameter of the model) the state of the system is replaced by white noise. 
Of course, this is a poor representation of the real experiment. 
However, we have found that it is often very suitable for a first qualitative picture of the effect of noise. 
Other simple models are the bit flip, phase flip, and amplitude damping channels.

A more precise description of the noise is possible with a Pauli channel~\cite{nielsen00}, 
where every tensor product of Pauli operators can appear as an error. 
These errors can occur with different probabilities.
A complete description of the error channel via Kraus operators~\cite{nielsen00} is also possible. 
The free parameters of this model can be determined via process tomography in the experiment~\cite{Mohseni2008,Flammia_2020}. 
It is a complex procedure that does not scale well with the system size but obtains complete information. 
It is worthwhile, for example, if one wants to obtain a very precise picture of a single gate of a quantum computer.
Instead of determining the parameters experimentally, one can also use "realistic" noise models. 
In this case, one tries to understand and model the physics of the process as well as possible.  
Often the parameters of the model have a physical interpretation. 
This approach is very hardware-specific and requires an exact fit of the model to the experiment. 
However, it also offers the chance to draw conclusions about necessary hardware improvements from model calculations, 
which is very helpful in the paradigm of hardware-software co-design.

We follow yet another approach in which the noisy process is largely considered as a black box, 
with few assumptions to be made about the noise~\cite{wimmer2023calibration}. 
In this study, the assumption of Pauli noise (see above) and the assumption that the noise of a circuit block is independent of the context. 
This means, that the same gate causes the same noise at different positions in the circuit. 
Of course, these assumptions might not be fully satisfied in a real experiment.
Instead of a description that is as complete as possible, we obtain information regarding errors that affect operators in the stabilizer elements,
which we will discuss below.

\subsubsection{Error Mitigation}
\label{sec:conceptual:errors:mitigate}

Error mitigation is a form of post-processing in which one tries to infer the ideal result from the noisy result~\cite{Cai2023quantum,endo2019a}. 
The techniques of error mitigation use additional measurements to extract information about the noise, which can be partially removed from the outcome.
They go beyond simply improving the measurement statistics by increasing the number of runs.
However, they are interesting for \NISQ computers because they do not require additional quantum resources.
Prominent examples are zero-noise extrapolation~\cite{Temme2017} and readout error mitigation~\cite{beisel2022}. 

Furthermore, the method of~\cite{wimmer2023calibration} to model the noise above can be used as an error mitigation technique. 
The parameters of the error model are determined via a calibration measurement. 
It allows us to infer the ideal expected values from the noisy ones of the stabilizer elements.
The results are comparable with readout error mitigation, while the method generates significantly less effort.

\subsubsection{Error Correction}
\label{sec:conceptual:errors:correct}

The topic of quantum error correction is vast and plays an important role. 
In this subsection, we briefly sketch the relevant concepts and refer interested readers to excellent introductions in~\cite{Gottesman2009introduction,bacon_2013,Roffe2019}. 
Generally speaking, it is the extension of classical error correction codes to correct not only bit-flip but also phase-flip errors, 
and thus also general errors on a quantum system.

Quantum error correction codes encode $k$ logical qubits into $n$ physical qubits.
Many codes can be described via the stabilizer of this code space, \ie, via a subgroup of the Pauli group whose elements leave the code words invariant. 
Small size examples are the 9-qubit Shor code~\cite{Shor1995}, the 7-qubit Steane code~\cite{Steane1996}, and the 5 qubit code~\cite{laflamme1996perfect,Bennett1996}. 
A family of widely used codes is the surface code~\cite{Kitaev_1997}. 
The layout of the qubits follows a lattice structure with the stabilizer generators acting locally, 
a fact that makes these codes a natural choice for hardware platforms with a matching architecture, \eg, those based on superconducting qubits.

The capacity of a code to correct errors is described by distance $d$, the minimum Hamming distance between two code words. 
The information about an error in the system is determined via the \textit{syndrome measurements}. 
Here, one measures a set of observables that yield enough information to inform the correction operation. 
This measurement result is called a \textit{syndrome}.

The concept of fault tolerance is crucial in quantum computing~\cite{Gottesman2009introduction}. 
It is possible, with the help of quantum error correction codes and clever design of circuits, 
to perform arbitrarily long calculations despite the noisy operations.
One can simply choose an arbitrarily large code, if it does not introduce too much noise due to the overhead of the additional operations, and the existing faults cannot propagate badly. 
It allows us to push the noise down to a desired level.
The required quality of operations that achieves this scaling is called the threshold of the error correction scheme~\cite{Shor1996ft}.
The additional complexity can be hidden in an abstract layer of the stack, \eg, when focusing on higher layers. 
It allows us to develop an ideal \QC without having to consider the additional complexity of error correction at all times.

%% file: content/conceptual/verification.tex
\subsection{Verification and Benchmarking}
\label{sec:verification}

Verification aims to ensure that software fulfills its requirements, \eg, that the output is correct under certain preconditions for given inputs. 
Similar to conventional non-deterministic software the stochastic nature of most quantum algorithms pose a challenge to verification. 
That is, the same inputs can produce different results due to the intrinsic properties of quantum measurements 
but also due to the high level of noise on near-term hardware, see \autoref{sec:conceptual:hardware}.
A typical requirement that needs to be verified is that one obtains a high-quality solution, 
\eg, a result close to the desired result, with a sufficiently high probability. 
Here we focus on static verification, while what is sometimes also referred to as dynamical verification is covered in \autoref{sec:conceptual:se:testing}.
The task of verification can be addressed from two sides: 
First, from a formal point of view, given “working” hardware we need a theoretical proof that the quantum algorithm is correct. 
Second, from a practical point of view, we need to ensure that the algorithm is implemented correctly in code for the classical and quantum parts of the program. 
Both parts need verification and the latter finally needs to be correctly translated into the executable circuit, see also \autoref{sec:conceptual:compiling}.
The verification of quantum algorithms gives rise to an interesting research question \cite{Gheorghiu2019,Wang2008verification, Amy2019}: 
When quantum computers outperform conventional computers, how can we ensure that the algorithm is correct? 

Benchmarking is the quantification of the performance of software and hardware.
Because at the current state of \QC the question is not yet how fast can we get a result but how good are the results that we get, 
benchmarking usually refers to assessing the quality of hardware components. 
So in contrast to conventional computer science we do not compare different software or hardware with metrics like time to solution, yet.
In this context benchmarks are standardized, technology-agnostic methods to evaluate quantum computers. 
The result of benchmarks are metrics for the performance of a specific device. 
They should be treated with caution, as they only cover single aspects of the machine, 
may struggle with the different hardware approaches, see also \autoref{sec:conceptual:hardware},
and only measure the current state of the technology - not its future perspective. 
Also note that the score is also affected by software, in particular the compilation. 
Any good benchmark should fulfill a number of requirements. 
It should be accepted by scientists and industry alike. 
The score of the benchmark is a number or a yes/no-answer. 
The metric allows for a meaningful interpretation, which goes beyond that specific benchmark test. 
It should not give an advantage to one specific technology by construction, but the values can and will be better for some technologies than others, of course. 
The benchmarks should be well-defined and easy to understand. 
They should be efficiently implementable, which poses a limitation on the information content in practice 
and therefore requires them to focus on specific aspects of the performance. 
They should be reproducible, which is a challenge given the non-deterministic character of quantum computers.
Finally they need to be scalable so they can be applied to small and larger quantum computers to enable tracking the development progress.

The following quantities and methods are typically considered in the context of benchmarking. 
\begin{itemize}
\item The fidelity of state preparation, single and two-qubit gates, and measurements. 
      These numbers measure how close the implemented operation and the target operation are.
\item Coherence times, which describe how long the coherence of a system, \ie, its ability to interfere, is conserved. 
      In particular the number of gate operations which can be performed in the coherence time indicates how long quantum computations can be.
\item Cross-talk, \eg, how strong idle qubits are affected by gates acting on other qubit. 
      Due to its non-local nature this noise can be difficult to handle.
\item Hardware connectivity, \ie, how many qubits can directly interact.
\item State and process tomography are methods that allow for a full characterization of a quantum state and a quantum operation, respectively~\cite{dariano2003quantum, Dariano04, Artiles05}.
\item Randomized benchmarking~\cite{Gaebler2012} is a method to find the average gate fidelity of an important subset of gates, the so-called Clifford gates. 
      It relies on the Gottesman-Knill-Theorem, which shows that circuits only consisting of such gates can be efficiently simulated on a conventional computer~\cite{nielsen00}. 
      This then allows to randomly insert gates into a circuit and efficiently invert their net effect. 
      Then the deviation from an identity operation is linked to the average fidelity of the gates. 
      One advantage of this method is that the metric is not affected by state preparation and measurement errors.
\end{itemize}

In order to perform useful reproducible benchmarks, we need to define a suite of standard problems, 
ideally reflecting interesting target applications (like SPEC benchmarks~\cite{SPEC} for different classical hardware) or, \eg, basic operations and algorithms (like LINPACK~\cite{LINPACK}).
Existing benchmark suites for \QC include SupermarQ~\cite{tomesh2022supermarq} and Arline~\cite{kharkov2022arline}. 
There are also benchmark suites tailored towards specific applications, \eg, fermionic quantum simulation~\cite{dallairedemers2020application}.
In addition, for the special case of comparing quantum hardware and software, it might be helpful to further specify some constraints 
how those problems should be solved as different approaches might not be comparable, \eg, hardcoding the solution is not a fair comparison.
This can be tricky to achieve in practice as different assumptions or prior knowledge about the problem are often used in different solution approaches.

For comparison with classical computers, there exists a wide range of possible implementations:
We can plug in a classical computer at almost any stage from the level of the original application problem, 
over a transformed formulation suitable for a quantum algorithm, to the actual operations for a specific hardware (at least for small problems or theoretical runtime considerations).
And even on a classical computer, software can be more or less optimized, which influences its runtime by several orders of magnitudes (see \eg~\cite{RoehrigZoellner2022} for an example).
So for actual benchmarking results, one needs to provide many additional details on all used implementations as well as on the hardware 
to actually allow an insightful comparison and interpretation.
We further suggest to define separate benchmarking suites to address specific questions in the future:
\begin{itemize}
    \item \textbf{quantum supremacy}: These benchmarks compare the fastest implementation on a quantum computer 
          with the fastest, elaborated, existing software for classical hardware for different key applications. 
    \item \textbf{near term practicability}: these benchmarks compare the (estimated) costs to solution for interesting algorithms (with input data from applications) 
          for different quantum platforms and for classical hardware. 
    \item \textbf{performance and correctness}: these benchmarks assess the accuracy/quality of solutions 
          obtained with different quantum hardware and software stacks for mathematical test problems with known solutions. 
\end{itemize}

%% file: content/technical.tex
\section{System Architecture and Implementation}
\label{sec:technical}

\begin{figure}[t!]
\centering
    \input{figures/technical_stack}
    \caption{A technical overview over the platform supporting quantum software developers.}
    \label{fig:technical-stack:overview}
\end{figure}
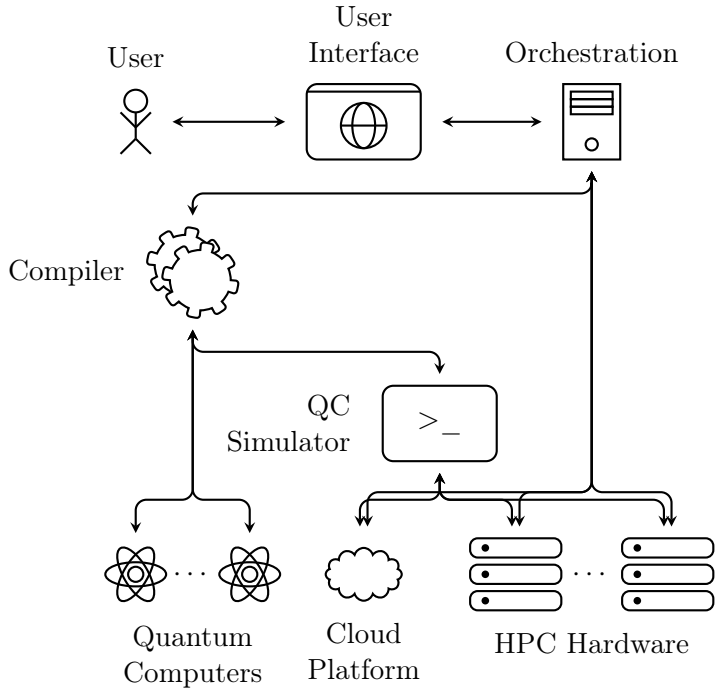

In \autoref{sec:conceptual}, we have described the conceptual workflow of quantum computers, from application to actual hardware.
The end users are typically ultimately interested in solving their engineering problem, \eg, in simulating the airflow around an aircraft or in finding some optimal resource scheduling.
To this end, we aim to construct a platform that allows end users to describe their domain-specific problem 
and find solutions to it while having to think about the underlying hardware as little as possible.
This section describes the technical building blocks we use to construct such a platform.
An illustration of the individual components and their connections is expressed in \autoref{fig:technical-stack:overview}.

We aim to construct this platform as domain-independently as possible.
To guide our description of the individual components, however, 
we use an artificial example problem from the automotive domain using this platform.
This example problem serves to highlight many of the considerations to be made when constructing a platform for quantum computing.
While other use cases will require additional considerations, 
we believe that this example already suffices to illustrate the most pressing and general concerns 
platform engineers should consider for a wide swath of use cases.

For our example, consider the goal of developing a new driving function for autonomous vehicles.
The engineers implementing this driving function want to evaluate whether it behaves safely in a number of specified driving scenarios.
To this end, they specify sets of possible scenarios using \TSC~\cite{DammMoehlmannPeikenkampEtAl2018}.
They then instantiate scenarios that conform to solutions of the given \TSC problem and simulate the behavior of the implemented driving function in that scenario~\cite{KroegerScheideggerBeckerEtAl2022}.
Moreover, they implement a software monitor that observes the simulation and reports unexpected behavior.
This is also accompanied by a visualization of the simulation. 
The full sequence of steps of the simulation shall be automatized.
To stay in the frame of a quantum software ecosystem, we further assume that the \TSC problem shall be solved using quantum computing hardware. 
This can, for instance, be done by converting the \TSC into a \SAT formula and use Grover's algorithm to search for feasible solutions provided by a corresponding quantum oracle gate.

Note that, in an actual application, the \TSC is converted into an \SMT formula instead of a \SAT formula, where the former is a strictly more general model than the latter.
There is, however, currently not a straightforward way to generate solutions for an \SMT formula using quantum circuits.
Hence, for the sake of example, we assume that the engineers instead generate concrete scenarios via \SAT formulas instead of \SMT formulas.

In practice, the described simulation steps require heterogeneous hardware:
The transformation from \TSC to \SAT and the extraction of a concrete scenario from a \SAT solution can be executed on virtually any hardware without proprietary software.
In contrast, the finding the solution of the \SAT problem and the execution of the simulation requires specialized software, 
namely \SAT solvers, such as Z3~\cite{MouraBjoerner2008}, and traffic simulation software, such as CARLA~\cite{DosovitskiyRosCodevillaEtAl2017}, respectively.
Moreover, the visualization requires specialized hardware, \eg, \GPUs.
As, in this running example, we assume that the engineer wants to find solutions to the \SAT formula using some quantum circuit, 
we also interact with quantum devices. 

In order to construct and execute their experiments, the engineers require some interface to the system.
This interface provides the engineer with an \IDE and, once the engineer is satisfied with their specification, passes the problem to some backend for execution.
We describe the requirements towards that interface in \autoref{sec:technical:interface}.
Once the end user has specified the problem, the platform will have to schedule the use of the heterogeneous hardware systems described above.
The major novelty of this platform lies in orchestrating the cooperation between classical computing hardware on the one hand, including, \eg classical workstations, \HPC resources, and \GPUs, and between \glspl*{qpu} on the other hand.
We describe the requirements towards this orchestration in \autoref{sec:technical:orchestration}.
The \glspl*{qpu} used during the execution may be implemented in actual hardware or it may be simulated using one of multiple quantum computing simulators.
We have described the constraints faced in using existing hardware platforms in \autoref{sec:conceptual:hardware}. 
The trade-offs to consider when using quantum simulators follows in \autoref{sec:technical:simulators}.

\subsection{User Interface}
\label{sec:technical:interface}

Quantum software developers require a straightforward interface for specifying their problems.
In our example, the end user must be able to specify the described loop consisting of reading a \TSC, calling external software, and performing computations on a well-suited \QPU.
The end user is not likely to be interested in the specifics of the underlying hardware but instead wants to have the choice of hardware handled by the platform during execution.
In contrast, the interface should also cater to experts who are not interested in specifying domain-specific problems but are working on developing novel quantum algorithms.
To this end, they require more direct access to the underlying hardware for, \eg, benchmarking.

The interface should allow the user to iterate rapidly on problem formulations \eg, the typical interface of \HPC hardware.
When using \HPC hardware for solving a domain problem, the underlying algorithms and implementations are often mature and well-tested.
In contrast, when using quantum computing hardware, the underlying algorithms and implementations are constantly evolving and are often adapted to the domain problem at hand.
Hence, the platform should allow the end user to rapidly iterate on the formulation of the domain-specific problem.

One approach to satisfy these requirements is allowing users to formulate their problems using a \textit{service-oriented architecture}.
In such an architecture, multiple independent software services collaborate to solve the specified problem.
In our example above, users could specify one service each for the following tasks: 
\begin{itemize}
  \item Transform a given \TSC into an \SAT formula,
  \item construct a quantum circuit that solves this formula,
  \item \label{serv:quantum-circuit} execute the quantum circuit to obtain a solution,
  \item transform the solution into a concrete scenario, and
  \item \label{serv:simulation} simulate and monitor the scenario using CARLA~\cite{DosovitskiyRosCodevillaEtAl2017}, obtaining a visualization of the simulation.
\end{itemize}
The user needs to specify the software and hardware requirements for each service, \eg, that they require a \QPU for the third service and CARLA with \GPUs for the visualization of the fifth service.
They do not necessarily have to implement all services themselves but can rely on other services that users of the platform have implemented and opted to share publicly.
Finally, the user must specify the data flow between these services and ask the orchestration component to execute the composed service.

Letting end users define composable services and publishing them to other users has proven successful in the context of data analysis with Apache Nifi~\cite{nifi} and in the context of preliminary design of airplanes, jet fuels, electrical grids, ships, and other complex systems with RCE~\cite{BodenFlinkFoerstEtAl2021}.
Moreover, a graphical user interface that allows users to graphically connect relevant services has been employed successfully for several decades in the field of data acquisition and analysis by LabVIEW~\cite{labview}.

\subsection{Orchestration and Data Management}
\label{sec:technical:orchestration}

Once the problem has been specified and is given to the orchestration component for execution, that component has to reserve computation time on the initial required computing resource.
In our running example, it requires some computation time on an off-the-shelf workstation which transforms the \TSC into an \SAT formula and subsequently transforms this formula into a quantum circuit.
The orchestration component then has to reserve computation time on some \QPU, either real hardware or simulated one to execute the quantum circuit.
Once the execution of the quantum circuit has finished and resulted in a solution to the \SAT problem, the orchestration component needs to reserve some computation time on an off-the-shelf workstation which transforms this solution into a scenario.
Subsequently, the orchestration component needs to reserve computation time on an \HPC resource equipped with \GPUs to simulate the generated scenario, monitor the simulation, and visualize the simulation if necessary.
Finally, the orchestration component needs to repeat the above steps until non-nominal behavior is observed during the simulation.

Our example shows that it is infeasible for the orchestration component to reserve all required computing resources prior to the execution of the initial service.
The requirements towards the \QPU, the available gates and number of qubits required for the execution of the quantum circuit, 
only become available after the execution of the initial service.
Hence, the orchestration component needs to be able to reserve computation time on the fly as results from earlier services become available.

Moreover, the orchestration component needs to take into account external requirements for the chosen computational resources.
The visualization of the simulation in the final step of the computation described above may require large maps, textures, or other large data artifacts to visualize the scenario with the required fidelity.
If these artifacts are only available to the visualization via a network connection with low bandwidth, the execution time of the complete computation will increase significantly.
Hence, the orchestration component needs to be aware of data-intensive parts of the computation and the locality of the required data.

\subsection{Use of QC Simulators}
\label{sec:technical:simulators}

In the section \autoref{sec:conceptual:hardware}, we have described the hardware platforms that are currently available for executing quantum circuits.
All these platforms are costly, only available in low quantities, do not provide a large number of qubits, and produce noisy results.
Although these problems are being addressed in the production of quantum computing hardware research, alternative solutions may tackle these issues.

A promising alternative is the use of quantum computing simulators that adopt classical hardware to simulate the execution of quantum circuits on actual hardware.
Simulating such an execution requires significant computing power that is usually only provided by \HPC systems.
These systems are typically the same ones that execute the classical part of the computation job.
Hence, any platform for the execution of quantum computing workloads using simulators must strike a balance between using the \HPC resources it has available for the simulation of quantum circuits and using them for classical computation.
Moreover, these \HPC resources are rarely available for exclusive use by the quantum computing platform.
Instead, the resources are also used for ``classical'' \HPC applications.
The owner of the resources has to balance their availability between the use by the platform and by the classical applications.

Although the results of the simulations produce data in the same order of magnitude as actual quantum computers (namely a few kilobytes or megabytes), they may offer additional diagnostic data which grows exponentially with the number of simulated qubits.
If this data is made available to end users, the platform needs to provide data storage as well as bandwidth for transferring the data to the end user.

%% file: figures/technical_stack.tex
\begin{tikzpicture}[thick]
    \node[label=above:User] (user) at (-3,0) {\usebox{\userbox}};
    \node[label={[align=center]above:User\\Interface}] (gui) at (0,0) {\usebox{\browserbox}};
    \node[label=above:Orchestration] (orchestration) at (3,0) {\usebox{\workstationbox}};
    
    \path[draw,thick,stealth-stealth] (user) -- (gui);
    \path[draw,thick,stealth-stealth] (gui) -- (orchestration);
    
    \node[label=left:Compiler] (compiler) at (-2.25,-2) {\usebox{\gearbox}};
    \node (qc-hardware-l) at (-3,-6) {\usebox{\quantumbox}};
    \node (qc-hardware-r) at (-1.5,-6) {\usebox{\quantumbox}};
    \node at ($(qc-hardware-l) ! .5 ! (qc-hardware-r)$) {$\cdots$};
    
    \node[anchor=north,align=center] at ($(qc-hardware-l.south) ! .5 ! (qc-hardware-r.south)$) {Quantum\\Computers};
    
    \node[label={[align=right]left:{QC\\Simulator}}] (simulator) at (1,-4) {\usebox{\programbox}};
    \node[label={[align=center]below:Cloud\\Platform}] (cloud) at (0,-6) {\usebox{\cloudbox}};
    
    \node[] (hpc-l) at (2,-6) {\usebox{\serverbox}};
    \node[] (hpc-r) at (4,-6) {\usebox{\serverbox}};
    \node at ($(hpc-l) ! .5 ! (hpc-r)$) {$\cdots$};
    
    \coordinate (qc-hardware-l-attach) at (qc-hardware-l.north|-hpc-l.north);
    \coordinate (qc-hardware-r-attach) at (qc-hardware-r.north|-hpc-l.north);
    
    \VerConnector{compiler.south}{simulator.north}
    
    \coordinate (cloud-attach) at (cloud.north|-hpc-l.north);
    \coordinate (cloud-1) at ($(cloud-attach) - (.05cm,0)$);
    \coordinate (cloud-2) at ($(cloud-attach) + (.05cm,0)$);
    \VerConnector{simulator.south}{cloud-1}

    \coordinate (hpc-l-1) at ($(hpc-l.north) - (.05cm,0)$);
    \coordinate (hpc-l-2) at ($(hpc-l.north) + (.05cm,0)$);
    \coordinate (hpc-r-1) at ($(hpc-r.north) - (.05cm,0)$);
    \coordinate (hpc-r-2) at ($(hpc-r.north) + (.05cm,0)$);
    \VerConnector{simulator.south}{hpc-l-1}
    \VerConnector{simulator.south}{hpc-r-1}
    
    \coordinate (interconnect-high) at ($(compiler.north) ! .5 ! (orchestration.south)$);
    \coordinate (simulator-attach) at (compiler.north-|simulator.north);
    
    \path[draw,thick,stealth-stealth,rounded corners] (orchestration.south) |- (interconnect-high) -| (compiler.north);
    
    \coordinate (hpc-middle) at ($(hpc-l.north) ! .5 ! (hpc-r.north)$);
    \coordinate (interconnect-low) at ($(hpc-r.north) ! .5 ! (simulator.south)$);
    \path[draw,thick,stealth-stealth,rounded corners] (orchestration.south) |- (interconnect-high-|hpc-middle) -- ($(hpc-middle|-interconnect-low) + (0,.1)$) -| (hpc-l-2);
    \path[draw,thick,stealth-stealth,rounded corners] (orchestration.south) |- (interconnect-high-|hpc-middle) -- ($(hpc-middle|-interconnect-low) + (0,.1)$) -| (hpc-r-2);
    \path[draw,thick,stealth-stealth,rounded corners] (orchestration.south) |- (interconnect-high-|hpc-middle) -- ($(hpc-middle|-interconnect-low) + (0,.1)$) -| (cloud-2);
    
    \path[draw,thick,stealth-stealth,rounded corners] (compiler.south) |- (interconnect-low-|qc-hardware-l) -- (qc-hardware-l);
    \path[draw,thick,stealth-stealth,rounded corners] (compiler.south) |- (interconnect-low-|qc-hardware-r) -- (qc-hardware-r);
    
    \node[anchor=north,align=center] at ($(hpc-l.south) ! .5 ! (hpc-r.south)$) {HPC Hardware};

\end{tikzpicture}

%% file: content/conclusion.tex
\section{Conclusion}
\label{sec:conclusion}

Quantum computing represents a paradigm shift in computational capabilities, with potential applications in various sectors. A key aspect to unlock its full potential is the establishment of a robust software ecosystem. This ecosystem not only provides the essential infrastructure for operating quantum devices but also serves as a bridge, enabling a broad spectrum of researchers, scientists, and industry experts to explore, use and enhance the applications of these quantum systems.

Our chapter takes a research-driven approach towards constructing such an ecosystem. We have bifurcated our exploration into two key dimensions. Firstly, the conceptual design which encompasses considerations from computational paradigms, applications like quantum simulation, over device optimized compiling to error handling, verification and benchmarking. This underscores the theoretical foundation, taking into account the unique challenges and attributes of quantum computing. Secondly, we delve into the system architecture and implementation, focusing on aspects ranging from user interfaces to orchestration, data management, and the critical role of quantum computing simulators. The fusion of these two perspectives ensures a comprehensive understanding and a holistic approach to developing a quantum software ecosystem.

As we step into the future, it is imperative to emphasize that this endeavor is iterative. Practical evaluation and real-world implementation of the proposed ecosystem will undoubtedly reveal areas of improvement. The scientific approach allows the adaptation of the ecosystem, especially given the rapidly evolving quantum hardware landscape. Monitoring these advancements and ensuring flexibility in the response will be critical to remaining aligned with the dynamic nature of quantum computing. By doing so, we pave the way for maximizing the potential of quantum computing, fostering innovation, and moving the field forward.


%% file: adds/acknowledgement.tex
The underlying research is part of the projects ‘Algorithms for quantum computer development in hardware-software codesign’ (ALQU), \url{https://qci.dlr.de/en/alqu},
and ‘Classical Integration of Quantum Computers’ (CLIQUE), \url{https://qci.dlr.de/en/clique}, 
which were made possible by the DLR Quantum Computing Initiative (QCI) and the German Federal Ministry for Economic Affairs and Climate Action (BMWK). 
Special thanks is due to IQM Quantum Computers and EleQtron for kindly making their respective quantum hardware chip designs available. 